\documentclass[fleqn,usenatbib]{mnras}

\usepackage{newtxtext,newtxmath}
\usepackage[T1]{fontenc}
\usepackage[dvipsnames]{xcolor}

\DeclareRobustCommand{\VAN}[3]{#2}
\let\VANthebibliography\thebibliography
\def\thebibliography{\DeclareRobustCommand{\VAN}[3]{##3}\VANthebibliography}

\usepackage{graphicx}	% Including figure files
\usepackage{amsmath}	% Advanced maths commands
\usepackage{hyperref}
\usepackage{multirow}
\usepackage{booktabs}
\usepackage{threeparttable}
\usepackage{subfigure}

\title[Superbubbles in simulated galaxies]{Evolution and distribution of superbubbles in simulated Milky Way-like galaxies}

\author[Chengzhe~Li et al.]
    {\parbox[T]{17cm}{Chengzhe Li$^{1}$\href{https://orcid.org/0000-0002-0480-2276}{\includegraphics[scale=0.8]{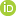}},
    Hui Li$^{1,2}$\thanks{E-mail: hliastro@tsinghua.edu.cn}\href{http://orcid.org/0000-0002-1253-2763}{\includegraphics[scale=0.8]{Images/orcid.png}},
    Wei Cui$^{1}$\thanks{E-mail: cui@tsinghua.edu.cn}\href{https://orcid.org/0000-0002-6324-5772}
    {\includegraphics[scale=0.8]{Images/orcid.png}},
    Federico Marinacci$^{3,4}$\href{https://orcid.org/0000-0003-3816-7028}
    {\includegraphics[scale=0.8]{Images/orcid.png}},
    Laura V. Sales$^{5}$\href{https://orcid.org/0000-0002-3790-720X}
    {\includegraphics[scale=0.8]{Images/orcid.png}}
    Mark Vogelsberger$^{6}$\href{https://orcid.org/0000-0001-8593-7692}
    {\includegraphics[scale=0.8]{Images/orcid.png}} and 
    Paul Torrey$^{7}$\href{https://orcid.org/0000-0002-5653-0786}
    {\includegraphics[scale=0.8]{Images/orcid.png}}\\
    }\\
     $^{1}$Department of Astronomy, Tsinghua University, Beijing 100084, China\\
     $^{2}$Department of Astronomy, Columbia University, New York, NY 10027, USA\\
     $^{3}$Department of Physics \& Astronomy ``Augusto Righi", University of Bologna, via Gobetti 93/2, 40129 Bologna, Italy \\
     $^{4}$INAF, Astrophysics and Space Science Observatory Bologna, Via P. Gobetti 93/3, I-40129 Bologna, Italy\\
     $^{5}$Department of Physics and Astronomy, University of California, Riverside, CA, 92521, USA\\
     $^{6}$Department of Physics \& Kavli Institute for Astrophysics and Space Research, Massachusetts Institute of Technology, Cambridge, MA 02139, USA\\
     $^{7}$Department of Astronomy, University of Florida, Gainesville, FL 32611, USA
    }

\date{Accepted XXX. Received YYY; in original form ZZZ}

\pubyear{2023}

\begin{document}
\label{firstpage}
\pagerange{\pageref{firstpage}--\pageref{lastpage}}
\maketitle

\begin{abstract}
Stellar feedback plays a crucial role in regulating baryon cycles of a galactic ecosystem, and may manifest itself in the formation of superbubbles in the interstellar medium. In this work, we used a set of high-resolution simulations to systematically study the properties and evolution of superbubbles in galactic environments. The simulations were based on the \textit{SMUGGLE} galaxy formation framework using the hydrodynamical moving-mesh code \textsc{Arepo}, reaching a spatial resolution of $\sim 4 \, \rm pc$ and mass resolution of $\sim 10^3 \, \rm M_{\sun}$. We identified superbubbles and tracked their time evolution using the parent stellar associations within the bubbles. 
The X-ray luminosity-size distribution of superbubbles in the fiducial run is largely consistent with the observations of nearby galaxies.
The size of superbubbles shows a double-peaked distribution, with the peaks attributed to
early feedback (radiative and stellar wind feedback) and supernova feedback. The early feedback tends to suppress the subsequent supernova feedback, and it is strongly influenced by star formation efficiency, which regulates the environmental density.
Our results show that the volume filling factor of hot gas ($T > 10^{5.5} ~\mathrm{K}$) is about $12 \%$ averaged over a region of 4 kpc in height and 20 kpc in radius centered on the disk of the galaxy.
Overall, the properties of superbubbles are sensitive to the choice of subgrid galaxy formation models and can, therefore, be used to constrain these models. 
\end{abstract}

\begin{keywords}
galaxies: evolution -- galaxies: ISM -- ISM: bubbles -- methods: numerical
\end{keywords}

%%%%%%%%%%%%%%%%%%%%%%%%%%%%%%%%%%%%%%%%%%%%%%%%%%

%%%%%%%%%%%%%%%%% BODY OF PAPER %%%%%%%%%%%%%%%%%%

\section{Introduction}

Stellar feedback plays a critical role in the evolution of galaxies. The absence of stellar feedback would result in excessive cooling of the interstellar gas, and the formed stellar mass would be much larger than the observed values \citep[i.e. the overcooling problem, see recent reviews, e.g.,][]{2017ARA&A..55...59N, 2020NatRP...2...42V}. Stellar feedback, mainly from massive stars, injects energy and momentum into the surrounding gas and regulates the phases of the interstellar medium (ISM). Among possible stellar feedback channels, supernovae and radiation are the most widely studied \citep[e.g.][]{2013MNRAS.428..129S, 2014MNRAS.439.2990S, 2016MNRAS.462.4532G}.
Stellar feedback can transport metal-rich gas to the circumgalactic medium (CGM) or even the intergalactic medium (IGM) through supernova driven galactic winds, and is therefore an important part of baryon cycling in a galaxy ecosystem \citep[][]{2017ARAA..55..389T}.

The most direct product of stellar feedback is the emergence of a multiphase ISM structure, with the hot phase mainly corresponding to superbubbles and galactic winds. Superbubbles are usually considered as large structures (about $100 ~\mathrm{pc}$ in scale) filled with hot gas that are blown by successive supernovae or stellar winds. 
The Solar System is located inside a superbubble \citep[i.e. the Local Bubble,][]{1987ARAA..25..303C, 1971ApJ...168L..33M}, which is mainly filled with highly ionized gas of temperature $T \sim 10^6 ~\mathrm{K}$ and electron density $n_{\rm{e}} \sim 0.005 ~\mathrm{cm^{-3}}$ \citep[][]{1998ApJ...493..715S}. Outside the Local Bubble lies cooler gas. The amount of neutral or partially ionized gas is expected to increase substantially on the boundary, which leads to a fairly accurate measurement of the Local Bubble extension, ranging from $100 \, \rm{pc}$ to $200 \, \rm{pc}$ for different sight lines \citep[e.g.][]{2003AA...411..447L, 2010AA...510A..54W}. Apart from the Local Bubble, other superbubbles were also discovered with X-ray observations in the Milky Way and nearby galaxies, such as the Large Magellanic Cloud (LMC). \citet{1990ApJ...365..510C} searched for diffuse X-ray emission around OB associations in the LMC and identified a number of superbubbles with high significance ($> 4 \sigma$). Fainter superbubbles were subsequently found \citep[][]{1995ApJ...450..157C,2001ApJS..136..119D}. Studies of other X-ray superbubbles were presented in \citet{1991ApJ...379..327W, 1994AA...283L..21B, 1996ApJ...467..666O, 1996rftu.proc..221B, 2003ApJ...599.1189C, 2003ApJ...590..306D, 2006AJ....131.2140T, 2007ApJ...656..928P, 2008ApJ...685..919T, 2009ApJ...707.1361T, 2012AA...547A..19K} and \citet{2018AA...619A.120K}, as well as of superbubble candidates in \citet{2010ApJS..187..495L, 2012AA...544A.144S, 2012ApJ...760...61S} and \citet{2020AA...637A..12K}.
The superbubbles that manage to break the cold gas disk of a galaxy would manifest themselves observationally as \ion{H}{I} holes, when viewed face-on \citep[e.g.][]{1986AA...169...14B, 2008A&A...490..555B,2020AJ....160...66P}. The size distribution of the observed \ion{H}{I} holes largely follows a power-law distribution with an index of about $-3$ \citep[][]{1997MNRAS.289..570O, 2011AJ....141...23B}.
Recently, several works on PHANGS galaxies have investigated some other superbubble-related structures in other wavelengths such as molecular shells and \ion{H}{II} regions \citep[e.g.][]{1_PHANGS, 2_PHANGS}.
Also, the Local Volume Mapper \citep[LVM][]{2019BAAS...51g.274K} is able to give a very detailed spectroscopic inspection of these structures by the integral-field spectroscopic survey.

Based on observations, \citet{1977ApJ...218..377W} constructed an analytical model for individual self-similar interstellar bubbles, powered by continuous injection of energy from stellar winds. This model was extended by \citet{1988ApJ...324..776M}, who showed that successive supernovae can also produce superbubbles and presented a very comprehensive scenario of the dynamical structure and evolution of superbubbles. In complement to observations and analytical models, numerical simulations have been carried out on stellar feedback processes \citep[e.g.][]{2004AA...425..899D, 2013MNRAS.432.1396G, 2014MNRAS.442.3013K, 2017ApJ...841..101L, 2017ApJ...834...25K, 2019MNRAS.489.4233M, 2021MNRAS.500.1833S, 2022MNRAS.510.5592A, 2023MNRAS.523.6336B}, and isolated superbubbles \citep[e.g.][]{2014ApJ...794L..21K, 2015ApJ...802...99K, 2017MNRAS.465.1720Y, 2018MNRAS.481.3325F, 2019MNRAS.490.1961E}. Simulations by \citet{2019MNRAS.490.1961E} suggested that both cooling at the interface and evaporation of shell gas into the interior of the bubble are important in the evolution of superbubbles. \citet{2017MNRAS.465.1720Y} performed three-dimensional simulations of isolated superbubbles and found that most of the input energy is lost by radiative cooling, and the superbubbles retain less than $10 \%$ of the input energy. Based on 1-d simulations, \citet{2020MNRAS.493.1034N} found that the size of superbubbles has a power-law distribution with an index of about $-2.7$.

Built upon the models for individual superbubbles, \citet{1977ApJ...218..148M} proposed a self-consistent ISM model (the so-called "three-phase model"), consisting of cold neutral medium ($T \sim 10^2 ~\mathrm{K}$), warm medium ($T \sim 10^4 ~\mathrm{K}$) and hot medium ($T \sim 10^6 ~\mathrm{K}$). The three-phase model is reasonably successful in explaining some of the important observations related to multi-phase ISM \citep[see][for a review]{1990ASPC...12....3M}. \citet{1974ApJ...189L.105C} suggested that a supernova event rate of about 1 per 50 years is sufficient for supernova remnants to overlap and form a network of "hot tunnels", with the hot gas filling about $10 \%$ of the interstellar volume. On the other hand, \citet{1977ApJ...218..148M} argued that the overall ISM structure would largely follow the local ISM, implying that the hot ISM occupies a large fraction of the Galactic disk and acts as a background phase. In their three-phase ISM model, the volume filling factor of hot gas can be larger than $70 \%$. In observations, it is difficult to directly measure the amount of hot gas in the Milky Way outside of the Local Bubble. Attempts were made to estimate the filling factor of hot ISM in nearby face-on galaxies \citep[][]{1996ApJ...468..102C}, but also turned out to be challenging, due to limited spatial and spectral resolutions.
Recently, simulations show that the filling factor of hot gas ranges from $17 \% \, - \, 44 \%$ in the ISM of galaxies \citep[][]{2004AA...425..899D}, depending on starburst activities.
However, previously with galactic-scale numerical simulations, no statistical analysis of superbubbles and their evolution was conducted. 

In this work, we study superbubbles systematically using state-of-the-art galaxy simulations. We are able to resolve individual superbubbles, track their evolution and perform a systematic study of superbubbles in high-resolution hydrodynamical simulations based on the \textit{SMUGGLE} model \citep[][]{2019MNRAS.489.4233M}. 
\textit{SMUGGLE} is a comprehensive, physically-motivated star formation and feedback model which can be widely used in detailed studies of galaxy evolution. \citet{2020MNRAS.499.5862L} presented a comprehensive study of giant molecular clouds (GMCs) in Milky-Way mass galaxies and found the influence of various model configurations (i.e. different feedback channels and different choices of star formation efficiency).
This model is also widely used in the studies of young massive clusters \citep[][]{2022MNRAS.514..265L}, black hole fueling \citep[][]{2022MNRAS.517.4752S}, Lyman-$\alpha$ properties \citep[][]{2022MNRAS.517....1S} and feedback in dwarf galaxies \citep[][]{2022PhRvL.129s1103B, 2022MNRAS.513.3458B, 2023MNRAS.520..461J}.

The outline of the paper is as follows. In Section~\ref{sec:methods}, we describe the ISM model adopted, the algorithms developed to identify superbubbles in 3-d maps and \ion{H}{I} holes in column density maps, and the methods used to find embedded stellar associations and track the evolution of superbubbles. In Section~\ref{sec:results}, the results about properties and evolution of superbubbles are presented. We discuss the implications of the results and open questions in Section~\ref{sec:discussion}, and briefly summarize the work in Section~\ref{sec:summary}.

\section{Methods}
\label{sec:methods} 

\begin{table}
	\centering
	\caption{Summary of the model configurations in the simulations we use. The "SN" ("Radiative  and wind") column tells whether the supernova (radiative and stellar wind) feedback is considered. $\epsilon_{\rm ff}$ is the star formation efficiency defined in Equation~(\ref{eq:sfr}).}
	\label{tab:config}
	\begin{tabular}{cccc} % four columns, alignment for each
		
		\hline
		Name & $\epsilon_{\rm ff}$ & SN & Radiative and wind \\
		\hline
		SFE1 & 0.01 & Yes & Yes \\
		SFE10 & 0.1 & Yes & Yes \\
		SFE100 & 1 & Yes & Yes \\
		SN & 0.01 & Yes & No \\
		Rad & 0.01 & No & Yes \\
		\hline
	\end{tabular}
\end{table}

We analyze a set of hydrodynamic simulations of isolated Milky Way-like galaxies presented in \citet{2020MNRAS.499.5862L}. The simulations were performed by the moving-mesh hydrodynamic code \textsc{Arepo} \citep[][]{2010MNRAS.401..791S} with a physically-motivated star formation and feedback sub-grid model, \textit{SMUGGLE} \citep[][]{2019MNRAS.489.4233M}. Star particles are formed in cold and dense molecular gas with a rate $\dot M_{\star}$ defined as
\begin{equation}
    \begin{aligned}
       \dot M_{\star} &= \epsilon_{\rm ff} \; \frac{M_{\rm gas}}{t_{\rm ff}},
	\label{eq:sfr}
	\end{aligned}
\end{equation}
where $\epsilon_{\rm ff}$ is the local star formation efficiency, $M_{\rm gas}$ is the mass of the gas cell, and $t_{\rm ff} = \sqrt{\frac{3 \pi}{32 G \rho_{\rm gas}}}$ is the free-fall time of the gas cell with density $\rho_{\rm gas}$.
The simulations reach a minimum softening length of $3.6 \, \rm pc$ and mass resolution of about $1.4 \times 10^3 \, \rm M_{\sun}$, which are thus able to spatially resolve the structure of superbubbles around stellar associations. To investigate the time evolution of superbubbles, we output the simulation snapshots every 1 Myr. We analyze simulations with different sub-grid models and parameters, as summarized in Table~\ref{tab:config}, to investigate how the properties of superbubble changes with galaxy formation physics. The SFE1 run is the fiducial run, which considers all stellar feedback channels with a star formation efficiency $\epsilon_{\rm ff} = 0.01$. The SFE10 and SFE100 runs change $\epsilon_{\rm ff}$ to 0.1 and 1, respectively.
The Rad run considers radiative and stellar wind feedback, while the SN run considers only supernova feedback. We refer the readers to \citet{2020MNRAS.499.5862L} for more details on the initial conditions and simulation setup. We identify and analyze the properties of superbubbles by applying the method described in the next subsection to the simulation snapshots between 0.25 Gyr and 1 Gyr.

\subsection{3D superbubble identification}
\label{sec:3d}

\begin{figure*}
    \centering
         \subfigure[SFE1]{
         \includegraphics[width=0.91\textwidth]{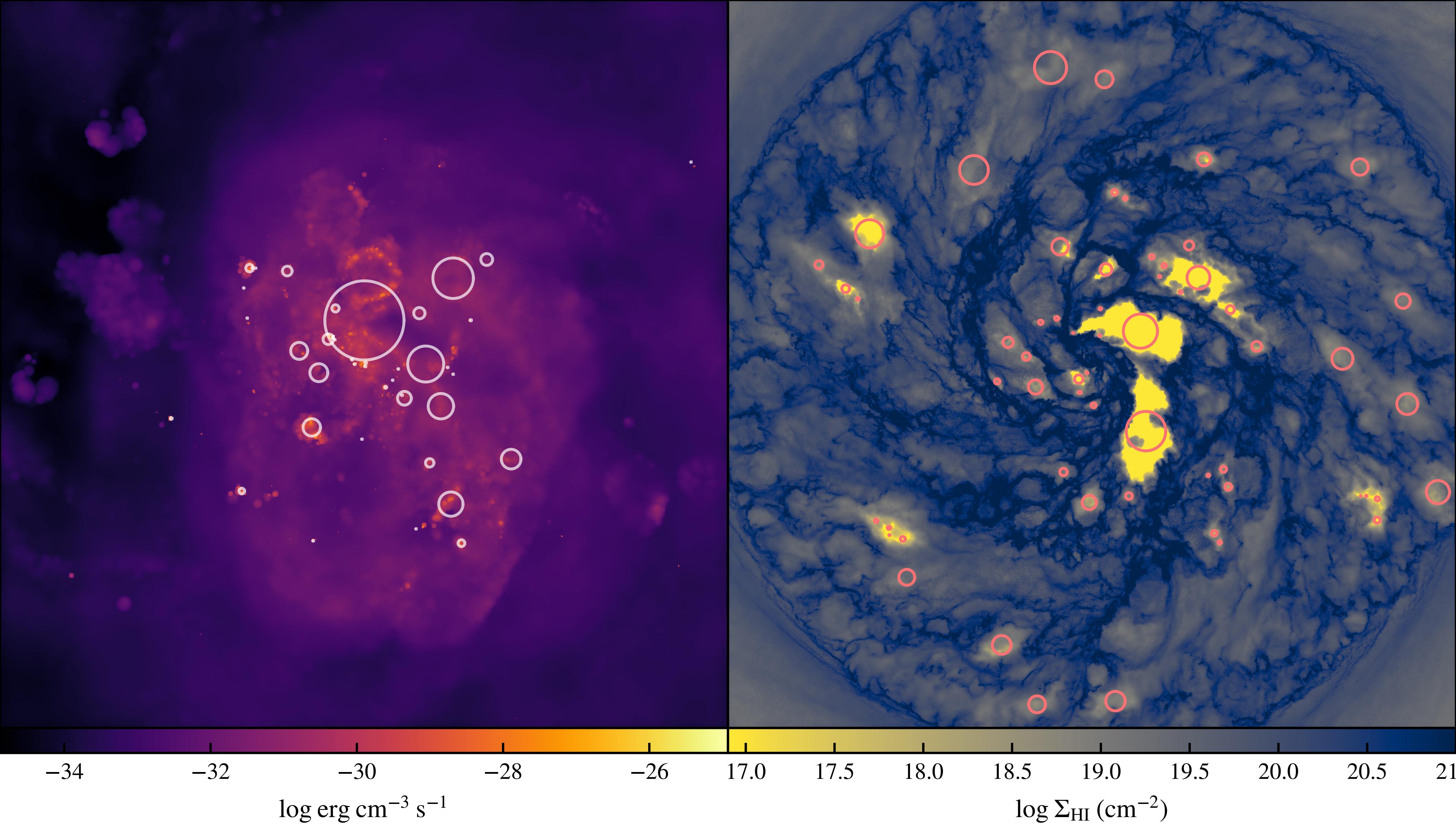}
         }
         \\
         \subfigure[SFE10]{
         \includegraphics[width=0.45\textwidth]{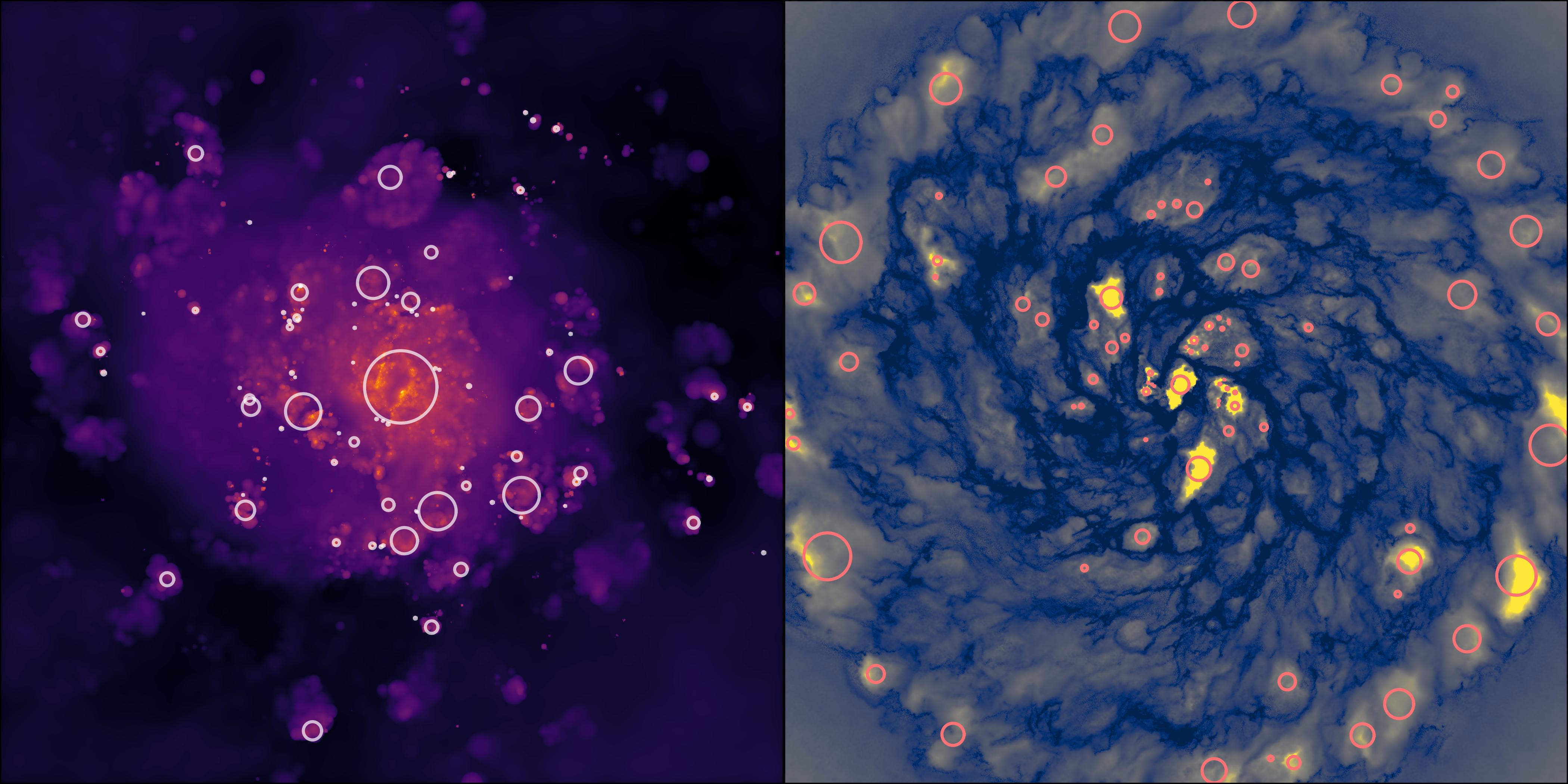}
         }
         \subfigure[SFE100]{
         \includegraphics[width=0.45\textwidth]{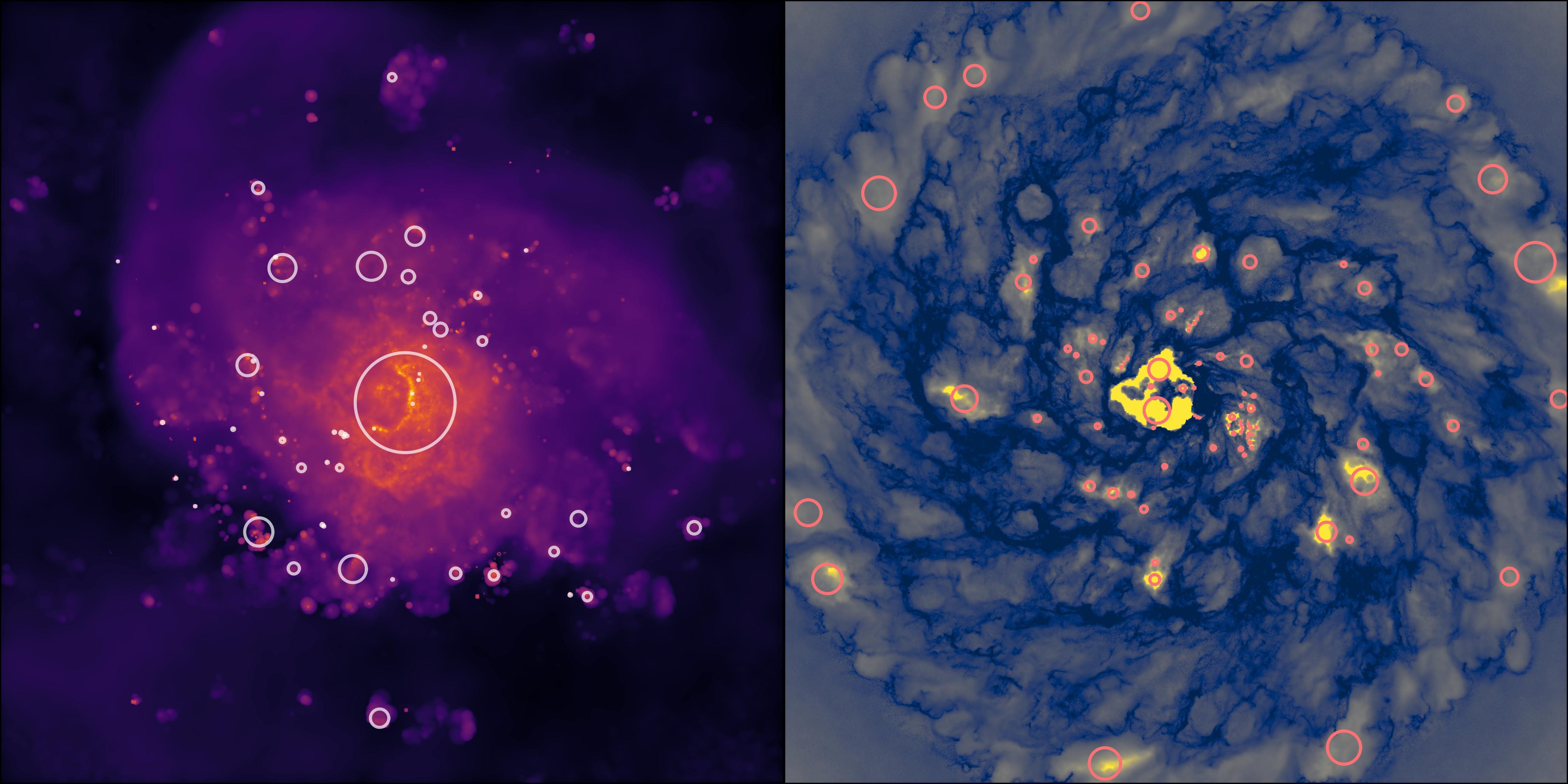}
         }
         \\
         \subfigure[SN]{
         \includegraphics[width=0.45\textwidth]{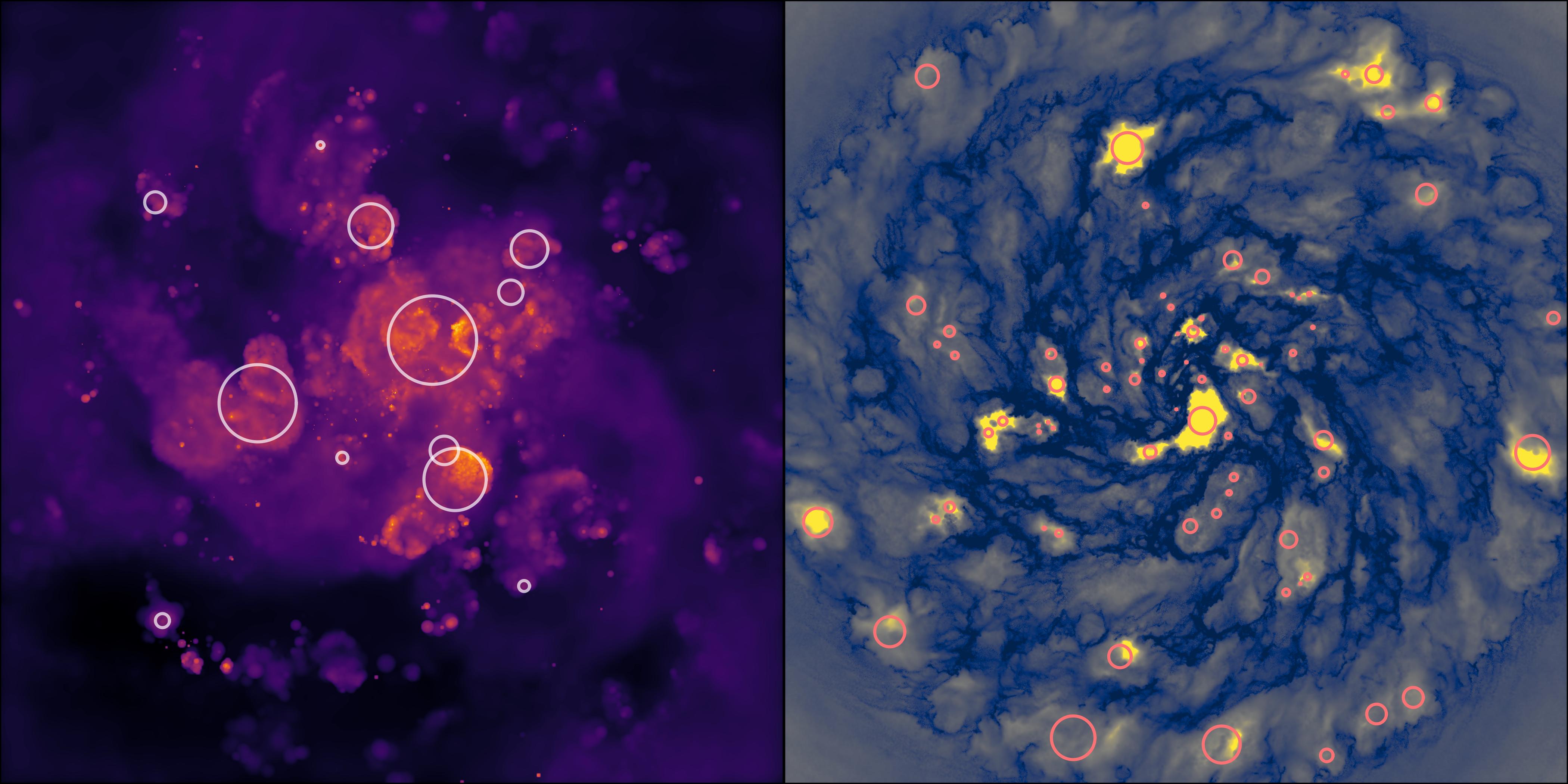}
         }
         \subfigure[Rad]{
         \includegraphics[width=0.45\textwidth]{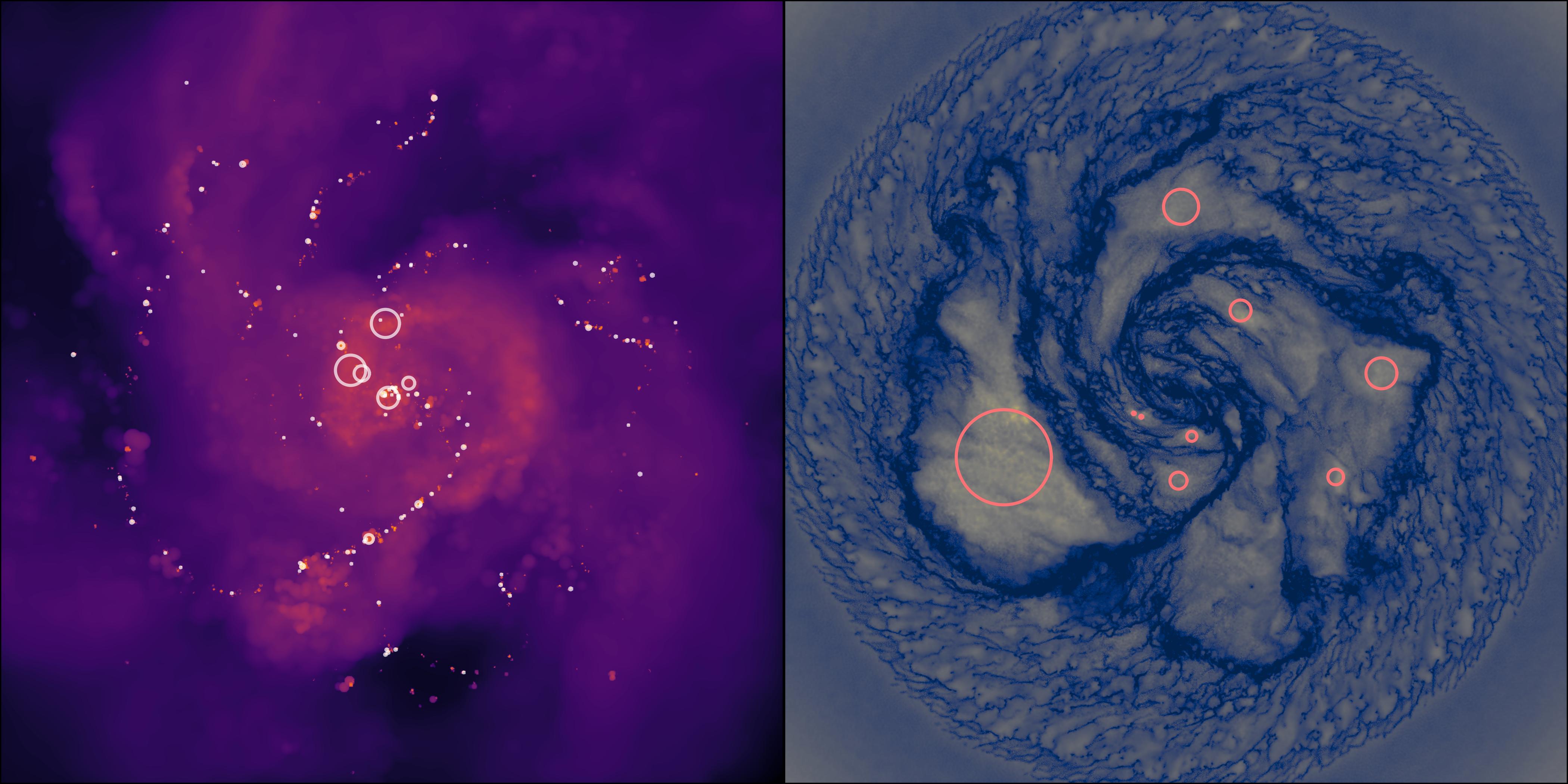}
         }
    \caption{\textit{Panel (a)}: Identification of superbubbles (left) and HI holes (right) in the SFE1 run. The left panel shows the distribution of X-ray emission (0.1 - 2.4 keV) at 0.4 Gyr for the SFE1 run in a face-on projection, with an extension of $40 \times 40 \; \rm kpc$. White circles correspond to the identified three-dimensional superbubbles, which are defined as hot, X-ray luminous structures. The positions and radii of the circles represent the projected location and the half-mass radii of the corresponding superbubbles. The right panel shows the face-on HI surface density map at 0.4 Gyr for the SFE1 run with an extension of $40 \times 40 \; \rm kpc$. Circles show the identified two-dimensional HI holes. \textit{Panel (b)-(e)}: similar to panel (a) but for SFE10, SFE100, SN and Rad run.}
    \label{fig:maps}
    \end{figure*}

Superbubbles are structures filled with hot gas blown out by stellar winds \citep[][]{1977ApJ...218..377W} or supernovae \citep[][]{1988ApJ...324..776M}. In the simulations, we consider superbubbles as structures of hot gas, and exclude structures that are outside the cold galactic disk and flow into the CGM. Superbubbles are born from stellar feedback in the cold ($T < 10^{3.5}$ K) disk. If the feedback process is strong enough, it may break the cold disk and blow out. Sometimes the blown-out gas may cut off the connection to the cold disk if feedback has already stopped or the cold disk is disturbed by some surrounding activities. We study hot gas within 2 kpc above and below the mid-plane, a region mainly composed of the cold disk and surrounding warm ($10^{3.5} < T < 10^{5.5}$ K) ionized gas. Beyond this region, the environment consists mostly of very diffuse hot gas, which can interfere with our superbubble finding algorithm. 

We first select gas cells within $2~\mathrm{kpc}$ above and below the mid-plane. We then find neighbours of each gas cell in order to prepare for the identification of structures.
In the framework of \textsc{Arepo}, the adaptive, dynamic Voronoi mesh is defined by the Voronoi tessellation of mesh-generating points. 
Therefore we only need to perform its topological dual Delaunay triangulation to the three-dimensional simulation output, and find the connected mesh-generating points to locate the neighbours of each Voronoi cell. After neighbours are found, we take gas cells with temperature $T > 10^{5.5} ~\mathrm{K}$ and density $\rho > 10^{-27} ~\mathrm{g / cm^{3}}$. This kind of gas is expected to efficiently emit X-rays, providing a convenient tracer for the identification of superbubbles.  Neighbouring gas cells satisfying the above selecting criteria will be assigned to a structure, which is considered as a superbubble candidate. 
Finally we need to exclude some unreliable structures to ensure our identification of superbubbles. We do not consider structures containing less than 5 gas cells or with radii smaller than $10~\mathrm{pc}$, and also exclude structures without stellar associations inside as they probably belong to some fragmented superbubbles or they are just parts of the hot, low-density background. 

For each identified superbubble, we define its center as the center of mass and its effective radius $r_{\rm{bubble}}$ as the half-mass radius of all the cells within. We have also tried using an ellipsoid to represent each superbubble. We estimate the semi-principal axes of the ellipsoid ($a$, $b$ and $c$ ) of each superbubble using the shape tensor \citep[e.g.][]{2011ApJS..197...30Z}, and calculate its effective radius as $R_{\rm ellipsoid} = (a b c) ^{1/3}$. It turns out that the results from these two methods agree well with each other, so we choose to use the half-mass radius to represent the size of superbubbles in this work for simplicity. On the left part of panel (a)-(e) in Fig.~\ref{fig:maps}, we show the location and radius of each identified superbubble over the the X-ray emissivity map in $0.1-2.4~\mathrm{keV}$ band from hot ($T > 10^{5.5} ~\mathrm{K}$) and optically thin ($n < 0.3 ~\mathrm{cm^{-3}}$) gas with the APEC model \citep[][]{2001ApJ...556L..91S}, which assumes collisional ionization equilibrium. The size of superbubbles can range from 10 pc to 1 kpc, with a typical size of about several hundred pc.

\subsection{Finding stellar associations located in superbubbles}
\label{sec:star_match}

In observations, the existence of OB associations offers important guidance to the identification of superbubbles \citep[e.g.][]{1990ApJ...365..510C}. Moreover, the properties of stellar associations, such as their mass, provide valuable constraints on the formation and evolution of the superbubbles. In our simulations, a stellar association is denoted as a star particle which represents a coeval simple stellar population \citep[][]{2019MNRAS.489.4233M} following the initial mass function presented by \citet{2001ApJ...554.1274C}. As gas is denoted by Voronoi cells, we match each star particle to the gas cell where it lies in, 
so we need to find the nearest mesh-generating point of that star particle.

We use the k-dimensional tree (KD-Tree) in SciPy \citep[][]{2020SciPy-NMeth} to find nearest points. We consider only stellar associations younger than $100$ Myr, as older stellar associations are not likely to contribute feedback any more. Some gas structures found in Section~\ref{sec:3d} may not contain stellar associations if they are parts of the hot background or belong to some fragmented superbubbles. In either case, we do not regard them as superbubbles. The mass of stellar associations have some important effects on the properties of superbubbles, and the related results are presented in Section~\ref{sec:stellar_mass}.

\subsection{Tracing the evolution of superbubbles}
\label{sec:tracking}

\begin{figure}
	\includegraphics[width=\columnwidth]{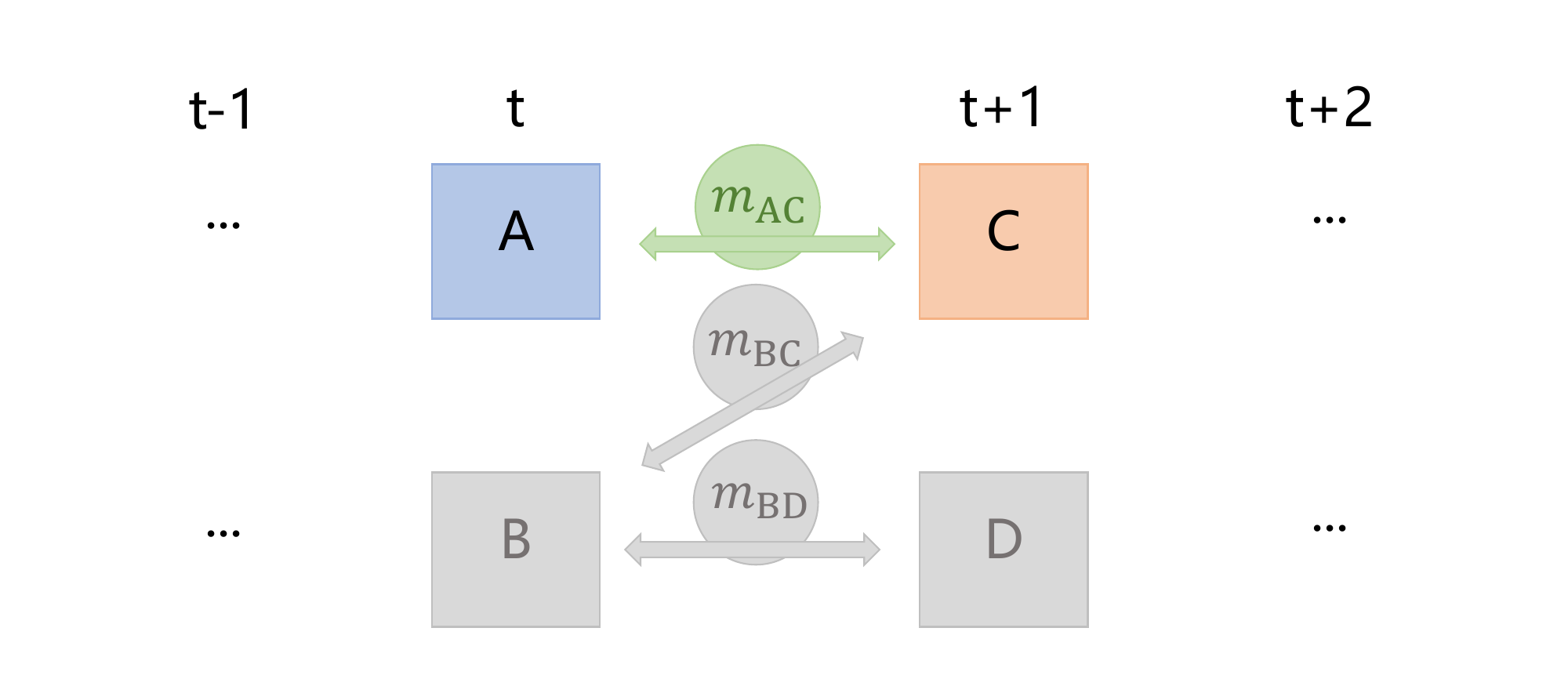}
    \caption{An example shows how we trace the superbubbles between snapshots. Superbubbles $A$ and $B$ are identified in snapshot $t$, while $C$ and $D$ are identified in snapshot $t + 1$. The total shared stellar mass between $A$ and $C$, $B$ and $C$, $B$ and $D$ are defined as $m_{\rm AC}, m_{\rm BC}$ and $m_{\rm BD}$. In this example we take $m_{\rm AC} > m_{\rm BC} > m_{\rm BD}$. $B$ is fragmented into two parts. One becomes $D$, and the other merges with $A$ into $C$.}
    \label{fig:link}
\end{figure}

Tracing the movement of gas is not trivial in a Voronoi mesh, where gas is always flowing across neighbouring Voronoi cells. Additionally, refinement and de-refinement will further complicate the analysis by removing or adding cells to the existing mesh.
 
Here we take the star particles in each superbubble as tracers. We assume that stellar associations will not have significant movements relative to the corresponding superbubbles, at least they will not totally leave the original superbubbles. This assumption seems quite reasonable and is also a basic assumption in observations to identify superbubbles, where most of the observed structures have their corresponding OB associations detected \citep[e.g.][]{1990ApJ...365..510C}.

In Section~\ref{sec:3d}, we have already identified superbubbles in each snapshot of the simulations, and what we need to do is to match the superbubbles with their progenitors and descendants in former and latter snapshots. The match follows a methodology similar to \citet{2012MNRAS.421.3488H}, but is not exactly the same, considering the features of our simulations and objectives. In general, several nearby superbubbles may merge into a bigger one, and when a big superbubble is dying, it may split into several smaller structures. For a superbubble in snapshot $t$, we define its progenitors (descendants) as superbubbles that share at least one star particle in snapshot $t-1$ ($t+1$). The main progenitor (descendant) of a superbubble is defined as the progenitor (descendant) containing the highest amount of shared stellar mass.
In snapshot $t$, a superbubble is born if there is no progenitor in snapshot $t-1$, and dies if there is no descendant in snapshot $t+1$. This approach allows us to establish connections among superbubbles across different snapshots, forming an evolutionary network with diverse evolutionary branches. 

Fig.~\ref{fig:link} shows an example, in snapshot $t$ we identify superbubbles $A$ and $B$, while in snapshot $t + 1$ we identify superbubbles $C$ and $D$. 
Superbubbles are linked with an arrow if they have some star particles in common. The total stellar mass in common between $A$ and $C$, $B$ and $C$, $B$ and $D$ are defined as $m_{\rm AC}, m_{\rm BC}$ and $m_{\rm BD}$. In this example we take $m_{\rm AC} > m_{\rm BC} > m_{\rm BD}$. 
Superbubble $B$ is fragmented into two parts. One becomes superbubble $D$, and the other merges with superbubble $A$ into superbubble $C$. Therefore, superbubble $C$ has two progenitors $A$ and $B$, while $A$ is its main progenitor. Superbubble $B$ has two descendants $C$ and $D$, while $C$ is its main descendant. As superbubble $A$ is the main progenitor of superbubble $C$ and also superbubble $C$ is the main descendant of superbubble $A$, the evolutionary branch from $A$ to $C$ is defined as a main branch.
We note that we do not consider superbubble $D$ as a newly born superbubble, because it is just a piece of the previously fragmented superbubble. Also, our main results are based on comparing the influence of different simulation configurations on superbubble properties. Using the identical analysis method on different simulation runs provides us a simple yet robust way of identifying superbubbles. Basic results on the size evolution of the main-branch superbubbles are shown in Section~\ref{sec:evolution}. A brief comparison to the analytical model will be discussed in Section~\ref{sec:analytic}.

\subsection{2D \ion{H}{I} hole identification}
\label{sec:2d}

Superbubbles triggered by feedback from their natal stellar associations can sometimes break the disk of cold gas above and below the mid-plane and puncture holes that are visible in \ion{H}{I} observations. In observations, we should note first that several studies rely on identification of \ion{H}{I} holes "by eye" \citep[e.g.][]{2011AJ....141...23B, 2020AJ....160...66P}, while there are some attempts trying to identify \ion{H}{I} holes automatically. For instance,  \citet{1998AA...332..429T} attempted identification using an expanding model, which is based on the superbubble model given by \citet{1977ApJ...218..377W}. \ion{H}{I} holes were identified with spatial and velocity information in the observed \ion{H}{I} data. The algorithm was refined in \citet{1999AA...343..352M} to consider simulation results of three-dimensional \ion{H}{I} shells. This kind of algorithms are only able to study structures with regular shapes, and do not work well when structures become non-spherical or incomplete.
A model independent hole-searching algorithm was developed by \citet{2005AA...437..101E}. It was first used to find high-velocity clouds in \citet{2002AA...391..159D}. The basic idea of this algorithm is to find local minima (or maxima) as the seeds of structures, then expand outward to take in brighter (or fainter) neighbour pixels. Based on similar ideas, here we adopt a tool, \textsc{astrodendro}, to identify \ion{H}{I} holes.

\textsc{astrodendro} \footnote{\url{http://www.dendrograms.org/}} \citep{2008ApJ...679.1338R} is a dendrogram-based algorithm, which is widely used in finding extended sources, such as giant molecular clouds (GMCs) and star-forming regions, for both observational \citep[e.g.][]{2022ApJ...929...74C} and simulations  \citep[e.g.][]{2020MNRAS.499.5862L} datasets. \textsc{astrodendro} turns the input map into a dendrogram, namely a tree that represents the hierarchy of the structures on the input map. 
The algorithm stops at a value $\sigma_{\rm base, min}$, which is mainly used to account for the background noise, and values below this threshold will not be considered. Another parameter $\sigma_{\rm delta, min}$ indicates a minimum significance for the structures in order to exclude "spikes" induced by fluctuations. The last parameter, $N_{\rm pix, min}$, represents a minimum pixel number of a structure to be identified. This is to avoid some spurious structures induced by the limitation of spatial resolution.

To identify \ion{H}{I} holes in simulation snapshots, we first generate the surface density map of the neutral hydrogen gas projected along the z-axis, namely a face-on projection of the gaseous disk. We create the \ion{H}{I} surface density map for the central $40 \times 40$ $\rm kpc^2$ region for simulations snapshots between 0.25 Gyr and 1 Gyr for all five runs. 
The map is resolved with $10000 \times 10000$ pixels, with the size of each pixel corresponding to 4~pc, which is approximately the limit of the adaptive softening of the simulations. 
\textsc{astrodendro} is developed to find peaks, but in our work we seek to find holes (valleys) in HI surface density ($\sigma_{\rm \ion{H}{I}}$) map. Therefore, we take the inverse of our HI surface density map so that HI holes become peaks in the inverse map ($\frac{1}{\sigma_{\rm \ion{H}{I}}}$). In practice, the value of HI surface density varies over six orders of magnitudes, so we feed the values of surface density map in logarithms ($- {\rm log} \,\sigma_{\rm \ion{H}{I}}$) into \textsc{astrodendro}. We set the threshold $\sigma_{\rm base, min} = - 19.5$ in the $- {\rm log} \,\sigma_{\rm \ion{H}{I}}$ map, which means a surface density upper limit of $10^{19.5} \, \rm cm^{-2}$ in the $\sigma_{\rm \ion{H}{I}}$ map. $\sigma_{\rm delta, min}$ is the minimum height difference between peaks, and it is also the minimum height of one peak to be recognized. Parameter $\sigma_{\rm delta, min} = 0.5 \; \rm dex$ in the $- {\rm log} \;\sigma_{\rm \ion{H}{I}}$ map means that we find peaks and their minimum heights are 0.5 dex. Therefore, in the $\sigma_{\rm \ion{H}{I}}$ map we find valleys (local minima), and their minimum heights are $10^{0.5} (\sim 3.162)$ times the local minima.
Finally, we take $N_{\rm pix, min} = 10$ in this work to avoid potential spurious structures.

The right part of panel (a)-(e) in Fig.~\ref{fig:maps} show the identified \ion{H}{I} holes. In general, \ion{H}{I} holes are located between the filamentary structures. Comparing to the left part of panel (a)-(e) of Fig.~\ref{fig:maps}, some \ion{H}{I} holes have co-spatial superbubble counterparts, but some do not. The relationship between superbubbles and \ion{H}{I} holes is complicated, and will be discussed in detail in Section~\ref{sec:bubble_vs_hole}.

\section{Results}
\label{sec:results}

\begin{table*}
	\centering
	\caption{Catalogue of X-ray identified superbubbles, organized by host galaxy. For each superbubble, we summarize its name, radius $r$, X-ray luminosity $L_{\rm X}$ and the corresponding OB association. References for each observed superbubble are also given.}
	\begin{threeparttable}
	\label{tab:catalogue}
	\begin{tabular}{ccccccc} % four columns, alignment for each
		\toprule[1pt]
		NO. & Name & $r$ (pc) & log $L_{\rm X}$ (erg / s) \tnote{${\rm a}$} & OB Association & Reference & Host Galaxy \\
		\toprule[0.6pt]
		\multirow{2}*{1} & \multirow{2}*{Local Bubble} & \multirow{2}*{165.0} & \multirow{2}*{36.08} & \multirow{2}*{UCL / LCC} \tnote{${\rm b}$} & \citet{1990ApJ...354..211S}; & \multirow{5}*{MW} \\
		 &  &  &  &  & \citet{2022Natur.601..334Z} & \\
		2 & Scorpius–Centaurus superbubble & 120.0 & 35.30 & USco \tnote{${\rm b}$} & \citet{2018AA...619A.120K} & \\
		3 & Cygnus superbubble & 225.0 & 36.70 (36.64) & Cygnus OB2 & \citet{1980ApJ...238L..71C} & \\
		4 & Orion-Eridanus superbubble & 113.0 & 37.61 & Orion OB1 & \citet{1996rftu.proc..221B} & \\
		\toprule[0.6pt]
		5 & N44 \tnote{${\rm c}$} \; (DEM 152 \tnote{$\rm d$} \;) & 27.4 & 36.15 & LH 47 \tnote{$\rm e$} \;, LH 48 & & \multirow{21}*{LMC} \\
		6 & N51D (DEM 192) & 53.0 & 35.52 & LH 51, LH 54 & \citet{1990ApJ...365..510C}; & \\
		7 & N57A (DEM 229) & 50.0 & 34.87 & LH 76 & \citet{1991ApJ...373..497W} & \\
		8 & N70 (DEM 301) & 55.0 & 35.26 & LH 114 & & \\
		\cmidrule{1-6}
		9 & N9 (DEM 31) & 54.5 & 35.00 (35.00) & LH 6 & \multirow{4}*{\citet{1995ApJ...450..157C}} & \\
		10 & N30A (DEM 105) & 63.2 & 35.28 (35.28) & LH 37 & & \\
		11 & N30C (DEM 106) & 34.2 & 34.95 (34.95) & LH 38 & & \\
		12 & DEM137 & 85.0 & 35.15 (35.15) & LH 43 & & \\
		\cmidrule{1-6}
		13 & N11 shell 1 (DEM 34) & 61.2 & 37.17 (37.90) & LH 9 & \multirow{9}*{\citet{2001ApJS..136..119D}} & \\
		14 & N103B (DEM 84) & 60.0 & 34.85 (35.00) & NGC 1850 & & \\
		15 & N105 (DEM 86) & 36.7 & 35.40 (35.62) & LH 31 &  & \\
		16 & N144 (DEM 199) & 47.4 & 34.74 (34.95) & LH 58 &  & \\
		17 & N154 (DEM 246) & 73.5 & 35.52 (35.75) & LH 81, LH 87 & & \\
		18 & N158 (DEM 269) & 54.8 & 35.85 (36.00) & LH 101, LH 104 & & \\
		19 & N160 (DEM 284) & 82.2 & 35.80 (35.95) & LH 103 &  & \\
		20 & N206 (DEM 221) & 45.0 & 35.09 (35.31) & LH 66, LH 69 & & \\
		21 & 30 Doradus C & 47.4 & 36.00 (36.19) & LH 90 & &  \\
		\cmidrule{1-6}
		22 & N185 (DEM 25) & 43.0 & 35.26 & - & \multirow{2}*{\citet{1996ApJ...467..666O}} & \\
		23 & N186 CE (DEM 50) & 50.0 & 35.62 & - &  & \\
		\cmidrule{1-6}
		24 & LMC-2 \tnote{$\rm f$} & 500.0 & 37.30 (37.33) & - & \citet{1991ApJ...379..327W} & \\
		\toprule[0.6pt]
		25 & NGC 604 & 122.5 & 36.16 (36.28) & - & \citet{2008ApJ...685..919T} & \multirow{2}*{M33} \\
		26 & IC 131 & 100.0 & 36.09 (36.43) & - & \citet{2009ApJ...707.1361T} & \\
		\bottomrule[1pt]
	\end{tabular}
	
	\begin{tablenotes}
	\footnotesize
	\item[*] We note that some superbubbles are identified in other wavelengths and do not have clear identifications in X-ray, such as the Ophiuchus superbubble \citep[][]{2007ApJ...656..928P} and the Perseus-Taurus superbubble \citep[][]{2021ApJ...919L...5B}.
        \item[${\rm a}$] The energy range used for the calculation of luminosities may vary between different observations. Corrected values for 0.1 - 2.4 keV band are shown in brackets for observations with available spectral fitting models.
	\item[${\rm b}$] There are three subgroups of the Scorpius–Centaurus association: Upper Centaurus Lupus (UCL), Lower Centaurus Crux (LCC) and Upper Scorpius (USco).
	\item[${\rm c}$] Following the name used in \citet{1956ApJS....2..315H}.
	\item[${\rm d}$] Named after \citet{1976MmRAS..81...89D}.
	\item[${\rm e}$] Named after \citet{1970AJ.....75..171L}.
	\item[${\rm f}$] Following the name used in \citet{1980MNRAS.192..365M}.
	\end{tablenotes}
	
    \end{threeparttable}
\end{table*}

\subsection{X-ray luminosity -- size distribution of superbubbles}
\label{sec:rad_lum}

\begin{figure}
	\includegraphics[width=\columnwidth]{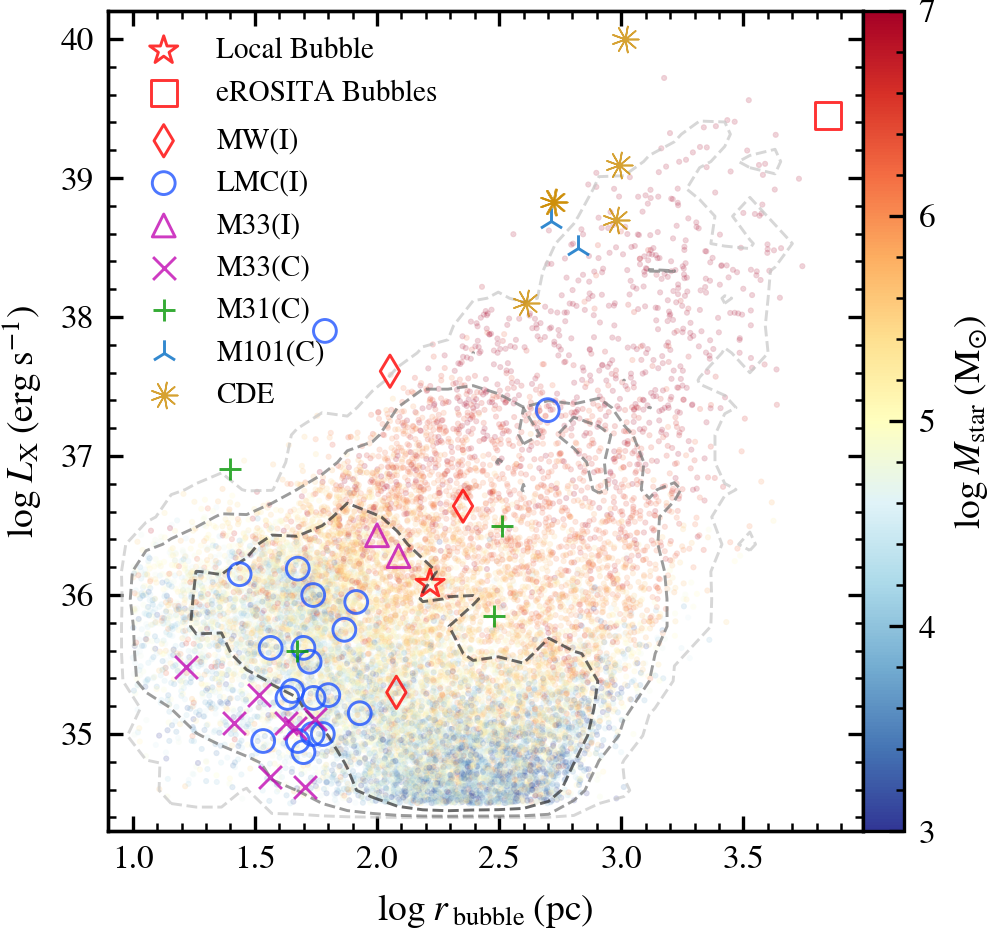}
    \caption{X-ray luminosity -- size distribution of superbubbles from simulations and observations. Each colored dot shows an identified superbubble in the SFE1 run, with its color representing the embedded stellar association mass. Dashed lines show the number density contours of 68\%, 95\% and 99.7\%. The Local Bubble and the eROSITA Bubble are denoted as red star and square. The X-ray identified superbubbles in the Milky Way, LMC and M33 are labeled with (I), while some candidates in M33, M31 and M101 with (C). A few examples of central diffuse emission (CDE) in other galaxies are shown for comparison.}
    \label{fig:rad_lum}
\end{figure}

X-ray observations provide direct information on superbubbles. Here, we summarize the key properties of the observed superbubble candidates in X-ray, and briefly compare them with those identified in our simulations. 
However, due to limitations in spatial resolution and sensitivity in the soft X-ray band, the observed sample size is relatively small, with over half of the identified superbubbles located in the LMC. In Table~\ref{tab:catalogue}, We have compiled the name, radius, luminosity of each superbubble identified with confidence in X-ray. Most of these bubbles have co-spatial OB associations detected. We should note that in some of the observations the superbubbles are considered elliptical in shape. In those cases, we compute the effective radius as $\sqrt{ab}$, where a and b are the semi-major and semi-minor axes of the ellipse.
Generally, the observed superbubbles have radii ranging from about 10 pc to several hundred pc and X-ray luminosities from $10^{34} ~\mathrm{erg \, s^{-1}}$ to $10^{37} ~\mathrm{erg \, s^{-1}}$.
It is important to note that these values were derived from data taken with different observatories, so the energy ranges used for computing luminosity may also be different. To address this issue, we have re-computed luminosity for the energy range of 0.1 - 2.4 keV, for cases in which spectral fitting results are available, and the results are shown in perentheses.
Additionally, different studies may adopt different algorithms to account for the foreground. These modelling uncertainties are expected to influence the luminosity values to some extent, but are unlikely to qualitatively affect the trend. 
Some superbubble-like candidates also exist in the nearby galaxies. For reference, we summarize the published results on those that were found in M31, M33 and M101 in Table~\ref{tab:candidates}.

As described in Section~\ref{sec:3d}, our identification of superbubbles is in principle able to find some kpc-scale giant bubbles breaking out from the galactic center, which have structures similar to the Fermi Bubbles \citep[][]{2010ApJ...724.1044S} and the eROSITA Bubbles \citep[][]{2020Natur.588..227P}. As mentioned earlier, such diffuse emissions can arise from stellar feedback or AGN feedback, or a combination of both. They can be classified as superbubbles if the central super massive black hole is not active. For comparison, Table~\ref{tab:central_diffuse} show the published results (likely incomplete) on the diffuse emission from the central region of a galaxy. These structures are usually large (exceeding $400 ~\mathrm{pc}$ in radius) and bright (with luminosity reaching $10^{38} ~\mathrm{erg \, s^{-1}}$ or higher in the 0.1 - 2.4 keV band), with large variation in size and luminosity, presumably due to their diverse galactic environment and intrinsic properties. 

Fig.~\ref{fig:rad_lum} shows a luminosity-size diagram for the superbubbles identified in the fiducial SFE1 run, along with identified X-ray superbubbles and candidates. We also show our compilation of galactic central diffuse emissions (CDEs, yellow asterisks). The Local Bubble and the eROSITA Bubbles are highlighted as red star and red square, respectively. Note that we calculate the X-ray luminosity of simulated superbubbles using the APEC model \citep[][]{2001ApJ...556L..91S} provided by AtomDB \citep[][]{2012ApJ...756..128F}, which assumes optically-thin gas, consequently we have excluded gas with density $n > 0.3 ~\mathrm{cm^{-3}}$, which lies in the optically-thick regime. For a better comparison with the observations, we specifically employ the half-luminosity radius, as opposed to the half-mass radii, to represent the size of superbubbles in Fig.~\ref{fig:rad_lum}. The identified superbubbles in our simulations vary greatly in size, spanning from 10 pc to several kpc. The luminosity $L_{\rm X}$ varies from $10^{34}$ to $10^{40} ~\mathrm{erg \, s^{-1}}$, showing a positive correlation with superbubble size and the mass of stellar association. Perhaps, larger mass of stellar association generally imply stronger feedback, which leads to superbubbles of larger size and higher luminosity. The X-ray luminosity-size distribution of superbubbles in our simulations is largely consistent with observations.

\subsection{Size distribution of superbubbles and \ion{H}{I} holes}
\label{sec:size}

\begin{figure}
	\includegraphics[width=\columnwidth]{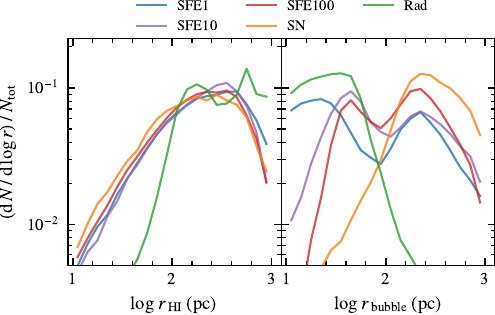}
    \caption{\textit{Left panel}: Size distribution of the identified 2D \ion{H}{I} holes in different runs. \textit{Right panel}: Size distribution of the identified 3D superbubbles in different runs.}
    \label{fig:size}
\end{figure}

As described in Section~\ref{sec:3d} and ~\ref{sec:2d}, we identify superbubbles from the 3D distribution of gas cells, and \ion{H}{I} holes from two-dimensional \ion{H}{I} maps. The size distribution of superbubbles is presented on the right panel of Fig.~\ref{fig:size}, showing complexity and variation across different runs.  Interestingly, in the Rad run, where supernova feedback is not considered, bubbles of hot gas can still form due to stellar winds and the radiation emitted by young massive stars and star clusters. When comparing the SN run to the Rad run, we find that superbubbles generated by supernova feedback tend to be much larger in size compared to those resulting from radiative feedback and stellar winds. Both the Rad run and the SN run exhibit a single-peak size distribution, with the Rad run peaking at tens of parsecs which represents the size of wind-blowing bubbles, and the SN run peaking at several hundred parsecs which represents the size of SN-driven bubbles. 
The remaining three runs exhibit double-peak distributions. Drawing on the findings from the Rad and SN runs, we infer that one is dominated by radiative and stellar wind feedback (at a smaller size), while the other peak by supernovae (at a larger size). By comparing SFE1, SFE10, and SFE100, we find that higher star formation efficiency results in a higher proportion of large superbubbles. We will see in Section~\ref{sec:stellar_mass} and Fig.~\ref{fig:stellar_mass}, that the larger the SFE, the larger the size of superbubbles dominated by early feedback, so more bubbles originated from early feedback will blend into the larger peak, making it appear like the larger peak is more prominent. The detailed impact of star formation efficiency will be discussed in Section~\ref{sec:SFE_fundamental}.

In principle, lower SFE should in general form smaller stellar associations, and should peak at a smaller size because of weaker feedback. However, as we will see from Section~\ref{sec:stellar_mass} that supernova-dominated (SN-dominated) superbubbles are not very sensitive to $\epsilon_{\rm ff}$. Moreover, in the simulations, the mass resolution of a star particle is $10^3 \; \rm M_{\odot}$. Star particle with $10^3 \; \rm M_{\odot}$ can generate about 6.5 SN in 10 Myr, and lead to a bubble with minimum radius of about 155 pc at 10 Myr for typical ambient density (see, e.g. the ambient density estimated in the SFE1 run as shown in Section~\ref{sec:SFE_fundamental}), so the place of larger peak might be limited by the resolution of star particle mass. With higher mass resolution, possibly we can see more SN-dominated superbubbles with smaller size.

The left panel of Fig.~\ref{fig:size} shows the size distributions of \ion{H}{I} holes in different simulation runs. The size distributions in runs with supernovae feedback are similar, ranging from $10 ~\mathrm{pc}$ to a kpc, with a peak at a radius of approximately $100 ~\mathrm{pc}$. However, in the Rad run, most of the identified \ion{H}{I} holes are considerably larger. This difference can also be seen from the \ion{H}{I} surface density maps in Fig. ~\ref{fig:maps}. They are probably low density voids between spiral arms/filaments and are not related to stellar feedback. It is similar to the observed HI voids in grand-design galaxies (like M51) \citep[e.g. ][]{2007A&A...461..143S}.
Early feedback alone is not powerful enough to frequently destroy the filaments and push gas to the inter-arm regions. This is also consistent with the two-point correlation function of giant molecular clouds analysed in \citet{2020MNRAS.499.5862L}.

The distribution of \ion{H}{I} hole size is very different from that of superbubbles, and shows only one peak in any run. The lack of the peak at smaller size might reflect the fact that \ion{H}{I} is not very sensitive to early feedback, because the processes are not as powerful as supernovae in pushing away the surrounding gas. Theoretically, the size distribution of \ion{H}{I} holes could be expressed as $N (R) \propto R^{1-2\beta}$ when assuming a power-law mechanical luminosity function $\phi (L) \propto L^{-\beta}$ for OB associations and a constant star formation rate \citep[][]{1997MNRAS.289..570O}. 
Typically in observations, the size distribution is fit with a power-law only for the larger sizes (i.e., $R_{\rm{hole}} > 100 ~\mathrm{pc}$), taking into account that the spatial resolution of observations may not be sufficient to resolve \ion{H}{I} holes smaller than $100 ~\mathrm{pc}$ \citep[][]{1997MNRAS.289..570O, 2011AJ....141...23B}. It is important to note that the larger size part does not seem to follow a simple power-law distribution in either simulations or observations. 
Furthermore, we find that the probability of finding \ion{H}{I} holes smaller than $100 ~\mathrm{pc}$ decreases even when the spatial resolution is sufficient to resolve them in our simulations. This tendency seems in line with the observations \citep[][]{1997MNRAS.289..570O, 2011AJ....141...23B}. The peak position is still limited by the resolution of star particle mass, as we mentioned above, even though the spatial resolution of the simulations is much higher than the observations.

\subsection{Effects of stellar association mass in superbubbles}
\label{sec:stellar_mass}

\begin{figure}
    \centering

        \includegraphics[width=\columnwidth]{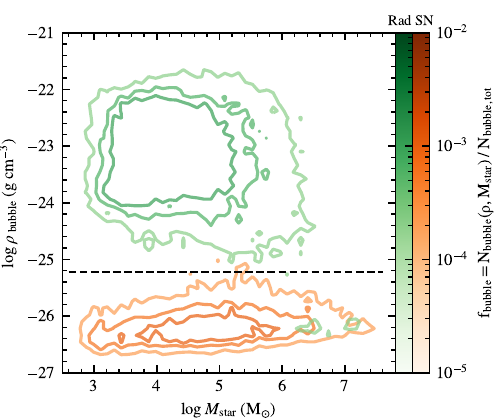}
         
    \caption{Distribution of superbubble density and stellar association mass for the SN (orange) and Rad (green) run. The horizontal dashed line shows the classification criterion $\rho_{\rm bubble} = 10^{-25.223} {\rm g \; cm^{-3}}$.}
    \label{fig:Mstar_dens_contour}
\end{figure}

\begin{figure}
    \includegraphics[width=\columnwidth]{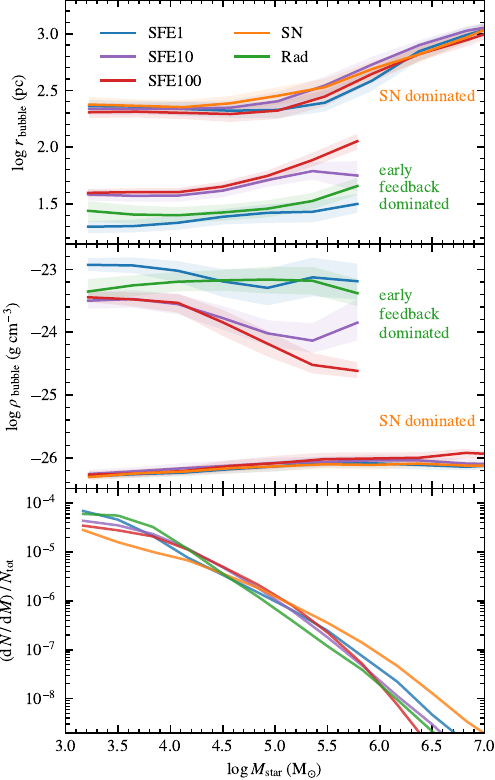}
    \caption{The effects of stellar association mass on the superbubble size (top panel) and density (middle panel). Two types of superbubbles (supernova-dominated and early feedback-dominated) are shown separately. Different runs are indicated with different colors. Solid lines show the median value of size and density at a particular mass of stellar association. Shaded regions corresponding to the $40^{\rm{} th} - 60^{\rm{} th}$ percentiles. The bottom panel shows the distribution of stellar association mass in different runs.}
    \label{fig:stellar_mass}
\end{figure}

A higher mass of stellar associations typically results in stronger stellar feedback processes that contribute to the formation of superbubbles. This factor is crucial in studying the properties of superbubbles in both simulations and observations. Moreover, the influence of stellar association mass reveals distinct variations across different runs, and can offer some valuable insights into the effects of star formation efficiency.

Firstly, the distribution of stellar association mass is shown as the bottom panel of Fig.~\ref{fig:stellar_mass}. The SN run has more massive stellar associations than other runs, but once early feedback is considered, the mass of stellar association is lowered. Therefore, early feedback plays a very important role on decreasing the mass of stellar associations, by suppressing the formation of star particles in the early stage, prior to supernova explosions. We will discuss more about this in Section~\ref{sec:interaction}.

In the SFE1, SFE10 and SFE100 run, there exist two types of superbubbles, which are dominated by supernova and early feedback. Fig.~\ref{fig:Mstar_dens_contour} shows the distribution of superbubble density and stellar association mass for SN and Rad runs. We can see that density of superbubble $\rho_{\rm bubble}$ is a pretty good property to classify these two types of superbubbles. We use logistic regression and find $\rho_{\rm bubble} = 10^{-25.223} {\rm g \; cm^{-3}}$ is a good criterion with 96\% accuracy. We applied this criterion onto all other runs and show in Fig.~\ref{fig:stellar_mass} the results on the effects of stellar association mass for these two types of superbubbles. We note that the classification of these two types of superbubbles is not perfect. In Fig.~\ref{fig:stellar_mass}, the larger the $\epsilon_{\rm ff}$, the larger the size of superbubbles dominated by early feedback, so more bubbles originated from early feedback will blend into the supernova-dominated type. 

The top panel of Fig.~\ref{fig:stellar_mass} illustrates the influence of stellar association mass, $M_{\rm{star}}$, on the size $r_{\rm{bubble}}$ of the superbubbles. For a particular stellar association mass, superbubbles dominated by supernova feedback tend to be larger than those dominated by early feedback. A positive correlation between $r_{\rm bubble}$ and $M_{\rm star}$ can be seen for early feedback-dominated superbubbles, and also for supernova-dominated superbubbles with stellar association mass larger than $10^5 \; {\rm M_{\odot}}$. However, the relation seems to become flat for supernova-dominated superbubbles when stellar association mass is lower than $10^5 \; {\rm M_{\odot}}$. This is possibly limited by the mass resolution of gas ($10^3 \; {\rm M_{\odot}}$) and the density criterion when classifying superbubbles (density lower than $10^{-25.223} {\rm g \; cm^{-3}}$). Together with the minimum cell number we set, which is 5, when identifying superbubbles in Section~\ref{sec:3d}, we can get a lower limit of radius as 178 pc (log 178 $\sim$ 2.25) for supernova-dominated superbubbles.

The middle panel of Fig.~\ref{fig:stellar_mass} provides information on the effects of stellar association mass on the density of superbubbles (estimated as the averaged density of its gas cells). Firstly, supernova-dominated superbubbles keep low densities (approximately $10^{-26} \, {\rm g \; cm^{-3}}$) across a wide range of stellar association masses, from $10^3 \, \mathrm{M_{\sun}}$ to $\mathrm{10^7 \, M_{\sun}}$. This may correspond to a typical density for overlapping supernovae. In contrast, early feedback-dominated superbubbles show significantly higher densities (above $10^{-24} \, {\rm g \; cm^{-3}}$), likely due to the relatively weaker dynamic effects of radiative and stellar wind feedback compared to supernovae. As stellar association mass increases, radiative and stellar wind feedback gain more strength to disperse the surrounding interstellar medium (ISM), resulting in lower superbubble densities. It is worth noting that the density $10^{-24} \, {\rm g \; cm^{-3}}$ largely corresponds to the typical density of the main interstellar medium (ISM) component, specifically phase (a) as shown in Fig.~\ref{fig:phase}. This indicates that early feedback-dominated superbubbles with lower stellar association masses primarily form thermally by the heating of photoionization, together with radiation pressure and stellar winds dispersing the original surrounding dense clouds to a density similar to the environment ISM density. Since early feedback-dominated superbubbles lack the significant power to sweep away the surrounding gas, they are primarily confined to the cold disk.

Fig.~\ref{fig:stellar_mass} shows that star formation efficiency $\epsilon_{\rm ff}$ does not have noticeable effect on supernova-dominated superbubbles, while superbubbles dominated by early feedback generally become larger and more diffuse when $\epsilon_{\rm ff}$ is higher. This indicates that $\epsilon_{\rm ff}$ mainly influences the processes of radiative feedback and stellar winds. In general, higher $\epsilon_{\rm ff}$ leads to lower environmental density. Radiative feedback is very sensitive to density $\rho_{\rm amb}$ and therefore this kind of superbubbles will be very different in size. Section \ref{sec:SFE_fundamental} will discuss this in more detail.

\subsection{Evolution of superbubbles}
\label{sec:evolution}

\begin{figure}
	\includegraphics[width=\columnwidth]{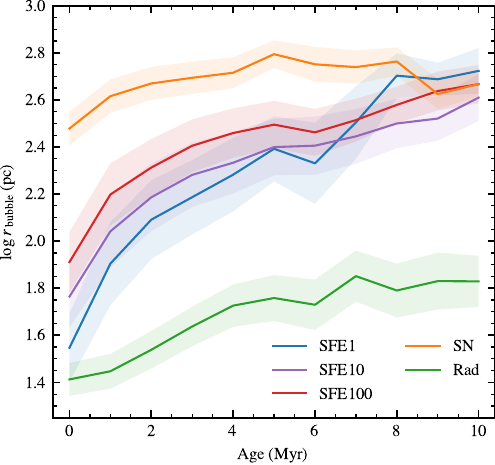}
    \caption{The evolution of superbubble size (solid lines) in different runs, which are shown in different colors, with a temporal resolution of 1 Myr. Shaded regions correspond to the $40^{\rm{} th} - 60^{\rm{} th}$ percentiles.}
    \label{fig:evolution}
\end{figure}

Based on the method described in Section~\ref{sec:tracking}, we use the star particles (each of which represents a stellar association) to link superbubbles in different snapshots and construct an evolutionary network including merging and splitting of superbubbles between snapshots.
In order to study the size evolution of superbubbles qualitatively, for simplicity, we consider only main branches of the evolutionary network. As defined in Section~\ref{sec:tracking}, the main progenitor (descendant) of a superbubble in snapshot $t$ is the superbubble containing the highest amount of shared stellar mass in snapshot $t-1$ (snapshot $t+1$). 
The evolution of main-branch superbubble size in each run is presented in Fig.~\ref{fig:evolution}, with a temporal resolution of 1 Myr. 

We can see that superbubbles in the SN run are larger than those in the other runs at the beginning. In the SN run, the lack of early feedback means the molecular clouds have a lot more time to grow larger. That will lead to more massive stellar associations and stronger feedback will happen on the birth of superbubbles.

Comparison across the SFE1, SFE10 and SFE100 runs reveals that star formation efficiency $\epsilon_{\rm ff}$ impact only the first few million years in the growth of superbubbles. This is consistent with the result shown in Section~\ref{sec:stellar_mass}, which is that $\epsilon_{\rm ff}$ mainly influences the processes in early feedback. In the beginning of superbubbles' life, they might be dominated by early feedback. Once supernova feedback starts, they may transit into supernova-dominated superbubbles, and therefore, the influence of $\epsilon_{\rm ff}$ gradually wanes. 
The possible transition from early feedback-dominated type to supernova-dominated type is also why it is difficult to separately study the evolution of these two types of superbubbles.
Further discussion about the effects of star formation efficiency is in Section~\ref{sec:SFE_fundamental}.

\subsection{Volume filling factor of the multi-phase gas}
\label{sec:filling_factor}

\begin{figure}
	\includegraphics[width=\columnwidth]{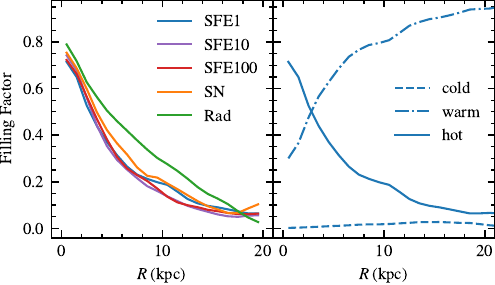}
    \caption{Spatial distribution of volume filling factor of different ISM phases. The interstellar gas is categorized into cold ($T < 10^{3.5}$ K), warm ($10^{3.5} ~\mathrm{K} < T < 10^{5.5}$ K) and hot ($T > 10^{5.5}$ K) phases. \textit{Left}: Radial profiles of hot gas volume filling factor in different runs. \textit{Right}: Radial filling factor profiles of cold (dash), warm (dashdot) and hot (solid) gas in the SFE1 run, with an extension to 20 kpc from the galactic center.}
    \label{fig:filling}
\end{figure}

In the SFE1 run, considering the gas within a cylinder centered at the galactic center, with a radius of $20 ~\mathrm{kpc}$ and a total height of $4 ~\mathrm{kpc}$, the overall volume filling factor of hot gas ($T > 10^{5.5} ~\mathrm{K}$) is found to be $12 \%$, which is similar to the findings of \citet{2004AA...425..899D}. In fact, the spatial distribution and filling factor of the multi-phase gas are not uniform. We divide the interstellar gas into cold ($T < 10^{3.5}$ K), warm ($10^{3.5} ~\mathrm{K} < T < 10^{5.5}$ K) and hot ($T > 10^{5.5}$ K) phases (a more detailed phase diagram is shown in Appendix \ref{sec:phase}). To investigate the radial variation of the filling factor, we maintain a constant thickness of $4$ kpc for the cylinder and divide it into multiple "rings" with different galactocentric radii. The non-uniform distribution of hot gas primarily arises from the non-uniform distribution of young stellar associations, as they are typically associated with stellar feedback.

The left panel of Fig.~\ref{fig:filling} shows the volume filling factors of hot gas in different "rings" and runs. The volume filling factor of hot gas peaks at $80 \%$ in the galactic center and gradually decreases with radial distance, without much variation across different runs. The right panel of Fig.~\ref{fig:filling} presents the radial profiles of the filling factors for different gas phases in the SFE1 run. Cold gas occupies only a small fraction of space, with hot gas predominantly filling the volume near the galaxy's center, while the warm gas phase dominates in outer regions.

\section{Discussion}
\label{sec:discussion}

\subsection{Relationship between superbubbles and \ion{H}{I} holes}
\label{sec:bubble_vs_hole}

Feedback from stellar associations result in a multi-phase structure of their surrounding environment. The enormous ejected energy and momentum give birth to hot, X-ray emitting superbubbles. The neutral gas in the vicinity is heated and ionized, leading to the formation of warm-gas shells \citep[which have been observed, e.g.,][]{2001ApJS..136..119D}, beyond which lies the neutral environment. Feedback with sufficient power can break through the disk of neutral gas above and below the mid-plane and puncture holes that are visible in the view of \ion{H}{I}. Therefore, one might envision the co-spatial existence of stellar associations, X-ray superbubbles and \ion{H}{I} holes. In reality, however, the conditions are much more complex.

The right part of panel (a)-(e) in Fig.~\ref{fig:maps} show the identified \ion{H}{I} holes. Comparing to the left part of panel (a)-(e) of Fig.~\ref{fig:maps}, we can see that some \ion{H}{I} holes have co-spatial superbubble counterparts, but some do not. This is not very surprising because they trace two different gas phases. Firstly, from Section~\ref{sec:size} we know that \ion{H}{I} gas is not to be sensitive to early feedback, therefore small superbubbles dominated by early feedback are not powerful enough to generate \ion{H}{I} holes. While stellar winds can in principle blow away the surrounding gas and reduce the density of \ion{H}{I}, it is less effective compared to supernova explosions. Moreover, an \ion{H}{I} hole may be destroyed by disturbances in the disk. On the other hand, superbubbles die when continuous stellar feedback ceases, and the hot gas subsequently cools down. After the death of a superbubble, the cavity blown out by stellar wind and supernovae may still remain there as a \ion{H}{I} hole. However, as the cavity loses pressure support from hot gas, it will easily shrink away or be disturbed by other surrounding activities. Finally, as mentioned in Section~\ref{sec:size}, some identified \ion{H}{I} holes in our simulations are probably just low density voids between spiral arms/filaments and are not related to stellar feedback.

\subsection{Effects of star formation efficiency}
\label{sec:SFE_fundamental}

Not all of the gas in giant molecular clouds will form stars in a free-fall time \citep[][]{2007ApJ...654..304K}. The star formation efficiency $\epsilon_{\rm ff}$, as defined in Equation~(\ref{eq:sfr}), determines the fraction of star forming gas and governs the local star formation rate. Based on the filling factor profiles in Section~\ref{sec:filling_factor}, we see that $\epsilon_{\rm ff}$ has minimal influence on the overall distribution of hot gas. However, properties of superbubbles show a strong dependence on $\epsilon_{\rm ff}$, as demonstrated in Section~\ref{sec:size}, \ref{sec:stellar_mass}, and \ref{sec:evolution}. There must be some fundamental effects of star formation efficiency. 

Superbubbles are basically formed through the interaction between stellar associations and the surrounding interstellar medium, so the properties of superbubbles will be regulated by both the environment and the energy injected by stellar feedback. For the SFE1, SFE10 and SFE100 runs, we estimate the average time interval between successive supernova events $\Delta t_{\rm SN}$ and the average density of the ambient medium $\rho_{\rm amb}$ (estimated as the average density within the range from $R_{\rm bubble}$ to $2R_{\rm bubble}$). We find their $\Delta t_{\rm SN}$ are 0.12, 0.15 and 0.12 Myr, and their $\rho_{\rm amb}$ are $1.51 \times 10^{-1}, \; 2.35 \times 10^{-2} \; {\rm and} \; 6.08 \times 10^{-3} \; \rm{} M_{\sun} \; pc^{-3}$.
This result indicates that simulation runs with different $\epsilon_{\rm ff}$ have similar supernova rates, but larger $\epsilon_{\rm ff}$ will lead to smaller density of the environment. This probably reveals differences in both the star forming stage and the early feedback stage. In the star forming stage, larger star formation efficiency means more efficient consumption of the cold gas, so the environment density will be lower. In the early feedback stage, radiative feedback and stellar winds heat the surrounding gas and reduce its density. Photoionization is very sensitive to the environment density $\rho_{\rm amb}$, as the Stromgren radius is proportional to $\rho_{\rm amb}^{-2/3}$. Consequently, the pre-processing of the surrounding gas becomes more pronounced.

\subsection{Interaction between early feedback and supernova feedback}
\label{sec:interaction}

While the SFE1 run considers both supernova feedback and early feedback, its results are not just simply the sum of the SN and Rad runs, and may give us some inspections about the interaction between early feedback and supernova feedback.

We note that early feedback, to some extent, suppresses the impact of supernova feedback.
Given the same star formation efficiency in the SN run and the SFE1 run, the SN run exhibits a greater abundance of huge superbubbles (with radius larger than $10^{2.4}$ pc) in Fig.~\ref{fig:size}. This can be attributed to the absence of early feedback, resulting in more massive stellar associations. As we see in the bottom panel of Fig.~\ref{fig:stellar_mass}, The SN run has more massive stellar associations than other runs, and  early feedback plays a very important role on decreasing the mass of stellar associations. For a given star-forming gas, early feedback starts at an early stage right after the formation of the first star particle. Therefore, it rapidly reduces the formation of star particles while gas clouds are still collapsing. Without early feedback, clouds keep collapsing and forming stars until the first SN explosion several million years later. The effect of early feedback on suppressing star cluster formation have also been shown in \citep[e.g.][]{2021MNRAS.506.3882S, 2024A&A...681A..28A}. Therefore, it consequently suppresses the power of supernova feedback and the formation of those huge superbubbles.

\subsection{Difference with observed X-ray luminosity - size distribution}
\label{sec:obs}

In section~\ref{sec:rad_lum} we have presented the X-ray luminosity - size distribution of superbubbles from our fiducial run as well as observations in Fig~\ref{fig:rad_lum}. While they are largely consistent, we note that the observed distribution seems to be a little biased towards smaller size. Firstly, the environment and history of various galaxies could be very different. Moreover, observations are actually limited by the sensitivity of instruments. For a superbubble at a given luminosity, larger in size means more diffuse and thus it has lower surface brightness and harder to be observed. As far as we learn from Fig~\ref{fig:rad_lum}, there appears to remain some faint superbubbles unexplored by current observations, so the sample size of observed superbubbles may hopefully increase in the future with advancements in more sensitive instruments.

\subsection{Comparison with analytical evolution model}
\label{sec:analytic}

\begin{figure}
	\includegraphics[width=\columnwidth]{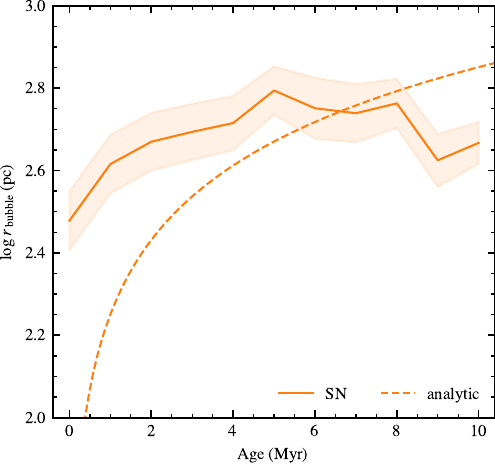}
    \caption{The evolution of superbubble size (solid line) in the SN run, with a temporal resolution of 1 Myr. Shaded region corresponds to the $40^{\rm{} th} - 60^{\rm{} th}$ percentiles. The dashed line represents the analytical evolution profiles for the SN run according to Equation~(\ref{eq:analytic_model}).}
    \label{fig:analytic}
\end{figure}

We compare the size evolution of superbubbles in the SN run with the analytical model given by \citet{1977ApJ...218..377W} and \citet{1988ApJ...324..776M}, in which self-similar superbubbles are generated by successive supernovae in a uniform environment. In the pressure-driven snowplow phase, the radius $R$ of a superbubble can be expressed as

\begin{equation}
    \begin{aligned}
        R &= \left (\frac{125}{154\pi} \right)^{1/5} L_{\rm SN}^{1/5} \; \rho_{\rm amb}^{-1/5} \; t^{3/5} \\
          &= 26.7 \; {\rm pc} \; \left( \frac{E_{\rm SN}}{10^{51} \rm erg} \right)^{1/5} \; \left( \frac{\Delta t_{\rm SN}}{\rm Myr} \right)^{-1/5} \; \left( \frac{\rho_{\rm amb}}{\rm M_{\sun} pc^{-3}} \right)^{-1/5} \; \left( \frac{t}{\rm Myr} \right)^{3/5}
	\label{eq:analytic_model}
	\end{aligned}
\end{equation}
where $t$ is the evolutionary time of a superbubble. The model has three parameters: the injected energy of each supernova event $E_{\rm SN}$ (it is set to $10^{51} \rm erg$); the average time interval between successive supernova events $\Delta t_{\rm SN}$; and the average density of the ambient medium $\rho_{\rm amb}$. To compare with the analytical model, we estimate $\Delta t_{\rm SN}$ and $\rho_{\rm amb}$ for each identified superbubbles. In the evolutionary network in our simulations, we record the number of supernova events in each time step and calculate the average time interval of supernovae. The average supernova rate $N_{\rm SN}$ in the SN run is about $25.3 \; \rm{} Myr^{-1}$, and we derive $\Delta t_{\rm SN}$ as the inverse of $N_{\rm SN}$, which gives $\Delta t_{\rm SN} = 0.04 \; \rm Myr$. $\rho_{\rm amb} = 1.91 \times 10^{-3} \; \rm{} M_{\sun} \; pc^{-3}$ is estimated as the average density within the range from $R_{\rm bubble}$ to $2R_{\rm bubble}$.

Substituting the parameters into Equation~(\ref{eq:analytic_model}), we can obtain predictions for the SN run using the analytical model, and the result is shown in Fig.~\ref{fig:evolution}. Evolution of superbubbles in the SN run is actually different from the analytical model. In the analytical model, it is assumed that supernovae explode one by one, with a uniform time interval $\Delta t_{\rm SN}$. However, we found that the number of supernovae is usually larger in the beginning and gradually decreases in our simulations. Moreover, while the analytical model assumes an unchanged uniform environment, in our simulations the expansion superbubbles will experience pressure confinement from the shocked surrounding materials, and this suppression is more significant as superbubbles grow larger.
We also note that the zero-age point we defined in our simulation is also different from that of the analytical model. In the analytical model, it is assumed that a superbubble is born immediately when stellar feedback begins. However, we define the birth of a superbubble at the time when we first found it in our simulations.

\subsection{Caveats}
\label{caveats}

Some caveats should be noted. First, the simulations were made for isolated $L_{*}$ galaxies. In comparison with observations, however, we included all superbubbles including those in dwarf galaxies like LMC and M33, due to the limited amount of X-ray observations. 
Second, the zoom-in simulations were made without considering the effects of inflows from the large scale structures into galaxies or AGN feedback. Although it is likely to lead to an inaccurate account of baryon cycles in galaxy ecosystems, it is not expected to significantly affect the properties of the ISM, which is the focus of this work. Contribution from ultraluminous X-ray sources \citep[ULXs,][]{2019AA...627A..63O,2023arXiv230200006P} is not considered as well.
Third, we have found that early feedback processes (radiative feedback and stellar winds) can impact the size and evolution of superbubbles. However, full radiative transfer is not implemented in the simulations; the effects of radiative feedback (including photoheating and radiation pressure) are calculated from an effective subgrid model \citep[][]{2010ApJ...709..308M}.
Fourth, the classification of two types of superbubbles in Section \ref{sec:stellar_mass} is not perfect, especially for high $\epsilon_{\rm ff}$ runs.
Fifth, in this work, superbubbles are identified as X-ray emitting bubbles, while the actual structures of superbubbles may be more complex. The cooler outskirts of superbubbles would emit at longer wavelengths.
Finally, we note that our automated \ion{H}{I} hole identifying method is not perfect and might include some local minima which are not related with feedback. However, the purpose of our study is to find the effects of different SFE and feedback processes on superbubbles and HI holes. The choice of parameters for HI hole identification is not expected to affect our main result that SFE does not have much impact on the distribution of HI hole size.

\section{Summary}
\label{sec:summary}

We use hydrodynamical simulations of isolated Milky Way-sized galaxies based on the SMUGGLE model in \textsc{Arepo} to study the properties and evolution of superbubbles. We are able to resolve and identify 2D \ion{H}{I} holes and 3D superbubbles, and construct evolution tracks of superbubbles using stellar associations as tracers. We conduct statistical analysis of superbubbles and take direct comparisons with both distribution in observations and evolution in analytical models. Here, we summarize our main results.

\begin{itemize}

\item The results from the fiducial run show a positive correlation between the X-ray luminosity and size of superbubbles. The correlation is in good agreement with observations of nearby galaxies.

\item The size distribution of \ion{H}{I} holes shows only one peak in all runs, and is not sensitive to radiative and stellar wind feedback. 

\item The superbubbles show a double-peaked size distribution. The size of superbubbles dominated by supernova feedback is generally larger than that of early feedback (radiative and stellar wind feedback) dominated superbubbles. The size distribution of superbubbles is different from that of \ion{H}{I} holes. In addition, not all superbubbles coincide spatially with \ion{H}{I} holes.

\item Star formation efficiency does not have significant influence on the overall distribution of hot gas but has significant impact on the properties of superbubbles. Larger star formation efficiency will lead to a higher proportion of large superbubbles with radius more than 100 pc. Star formation efficiency mainly influences the processes in early feedback, by regulating the environmental density. 

\item Early feedback plays an important role in suppressing supernova feedback and the mass of stellar associations. Through the process of heating and ionizing the surrounding environment of newly born stars, early stellar feedback actively restrains the formation of additional stars. Consequently, fewer star particles are formed compared to the scenario without early feedback, resulting in a decreased number of supernovae and thus a reduced supernova feedback.

\item The average volume filling factor of the hot gas is about 12 \%, but its distribution is not uniform across the galaxy. The filling factor reaches more than 80 \% at the galactic center and decreases towards the outer disk, where the warm gas mainly lies. The filling factor of the cold gas is a very low throughout the galaxy.

\end{itemize}

\section*{Acknowledgements}

We thank the anonymous referee for the suggestions that have significantly improved the manuscript, Volker Springel for allowing us access to the Arepo code, and Yu-Ning Zhang, Rui Huang and Jiejia Liu for very useful discussion and comments. This work was supported in part by the National Natural Science Foundation of China through Grant No. 11821303 \& 12373006, by the Ministry of Science and Technology of China through Grant 2018YFA0404502 and 2023YFB3002502, and by China Manned Space Program through its Space Application System. We also acknowledge the Tsinghua Astrophysics High-Performance Computing (TAHPC) platform for providing computational and data storage resources that have contributed to this work. The identification of 2D \ion{H}{I} holes in Section~\ref{sec:2d} made use of {\sc astrodendro}, a Python package to compute dendrograms of Astronomical data \footnote{\url{http://www.dendrograms.org/}}. LVS acknowledges financial support from grants  NSF-CAREER-1945310 and NSF AST2107993.

\section*{Data Availability}

The data that support the findings of this study are available from the corresponding author, upon reasonable request.



\bibliographystyle{mnras}
\bibliography{superbubbles} % if your bibtex file is called example.bib

%\begin{thebibliography}{99}
%\end{thebibliography}

%%%%%%%%%%%%%%%%%%%%%%%%%%%%%%%%%%%%%%%%%%%%%%%%%%

%%%%%%%%%%%%%%%%% APPENDICES %%%%%%%%%%%%%%%%%%%%%

\appendix

\section{Catalogue of superbubble candidates and central diffuse emissions}
\label{sec:candidates}

\begin{table}
	\centering
	\caption{Superbubble candidates in M33, M31 and M101. Corrected luminosities for 0.1 - 2.4 keV band are shown in brackets.}
	\label{tab:candidates}
	\begin{tabular}{ccccc} % four columns, alignment for each
		\hline
		NO. & Name & $r$ & log $L_{\rm X}$ & Reference\\
		 & & (pc) & (erg / s) & \\
		\hline
		\multicolumn{5}{|c|}{M33} \\
		1 & L10-012 & 26.0 & 35.08 & \citet{2010ApJS..187..495L} \\
		2 & L10-026 & 36.5 & 34.69 & \citet{2010ApJS..187..495L} \\
		3 & L10-043 & 42.5 & 35.08 & \citet{2010ApJS..187..495L} \\
		4 & L10-089 & 46.0 & 35.04 & \citet{2010ApJS..187..495L} \\
		5 & L10-092 & 50.5 & 34.62 & \citet{2010ApJS..187..495L} \\
		6 & L10-117 & 33.0 & 35.28 & \citet{2010ApJS..187..495L} \\
		7 & L10-119 & 16.5 & 35.48 & \citet{2010ApJS..187..495L} \\
		8 & L10-122 & 55.5 & 35.11 & \citet{2010ApJS..187..495L} \\
		\hline
		\multicolumn{5}{|c|}{M31} \\
		9 & K20 region 2 & 324.0 & 36.43 (36.50) & \citet{2020AA...637A..12K} \\
		10 & K20 region 3 & 300.0 & 35.76 (35.85) & \citet{2020AA...637A..12K} \\
		11 & SPH11-263 & 47.0 & 35.60 & \citet{2012AA...544A.144S} \\
		12 & SPH11-1234 & 25.0 & 36.90 (36.91) & \citet{2012AA...544A.144S} \\
		\hline
		\multicolumn{5}{|c|}{M101} \\
		13 & NGC 5461 & 516.2 & 38.46 (38.69) & \citet{2012ApJ...760...61S} \\
		14 & NGC 5462 & 663.3 & 38.23 (38.50) & \citet{2012ApJ...760...61S} \\
		\hline
    \end{tabular}
\end{table}

\begin{table}
	\centering
	\caption{Diffuse emission from central region of galaxies. Corrected luminosities for 0.1 - 2.4 keV band are shown in brackets.}
	\label{tab:central_diffuse}
	\begin{tabular}{ccccc} % four columns, alignment for each
		\hline
		NO. & Name & $r$ & log $L_{\rm X}$ & Reference\\
		 & & (pc) & (erg / s) & \\
		\hline
		\multirow{2}*{1} & eROSITA & \multirow{2}*{7000.0} & \multirow{2}*{39.01 (39.45)} & \multirow{2}*{\citet{2020Natur.588..227P}} \\
		 & Bubbles & & & \\
		2 & NGC 3079 & 1030.8 & 39.99 (39.99) & \citet{1998AA...340..351P} \\
		3 & IC 10 & 408.0 & 37.90 (38.10) & \citet{2005MNRAS.362.1065W} \\
		4 & NGC 1569 & 948.7 & 38.70 (38.70) & \citet{1995ApJ...448...98H} \\
		5 & NGC 4449 & 979.8 & 38.96 (39.09) & \citet{2003MNRAS.342..690S} \\
		6 & NGC 5253 & 530.0 & 38.60 (38.83) & \citet{2004MNRAS.351....1S} \\
		7 & NGC 4214 & 524.4 & 38.48 (38.83) & \citet{2004MNRAS.348..406H} \\
		\hline
	\end{tabular}
\end{table}

We list in Table~\ref{tab:candidates} the information of superbubble candidates in some nearby galaxies (M33, M31 and M101). They are not identified with confidence, but their size and luminosity are still valuable due to the very small sample size of confirmed superbubbles. They are also valuable targets for future observations and maybe some of these candidates can be confirmed as superbubbles in the future. In principle, our superbubble identification method is able to find some kpc-scale structures which are just breaking out from the galactic center. This kind of source is usually very extensive and bright. Table~\ref{tab:central_diffuse} gives some examples of central diffuse emissions (CDEs) from the Milky Way (i.e. the eROSITA Bubbles) and other galaxies.

\section{ISM-CGM phase structure}
\label{sec:phase}

\begin{figure}
	\includegraphics[width=\columnwidth]{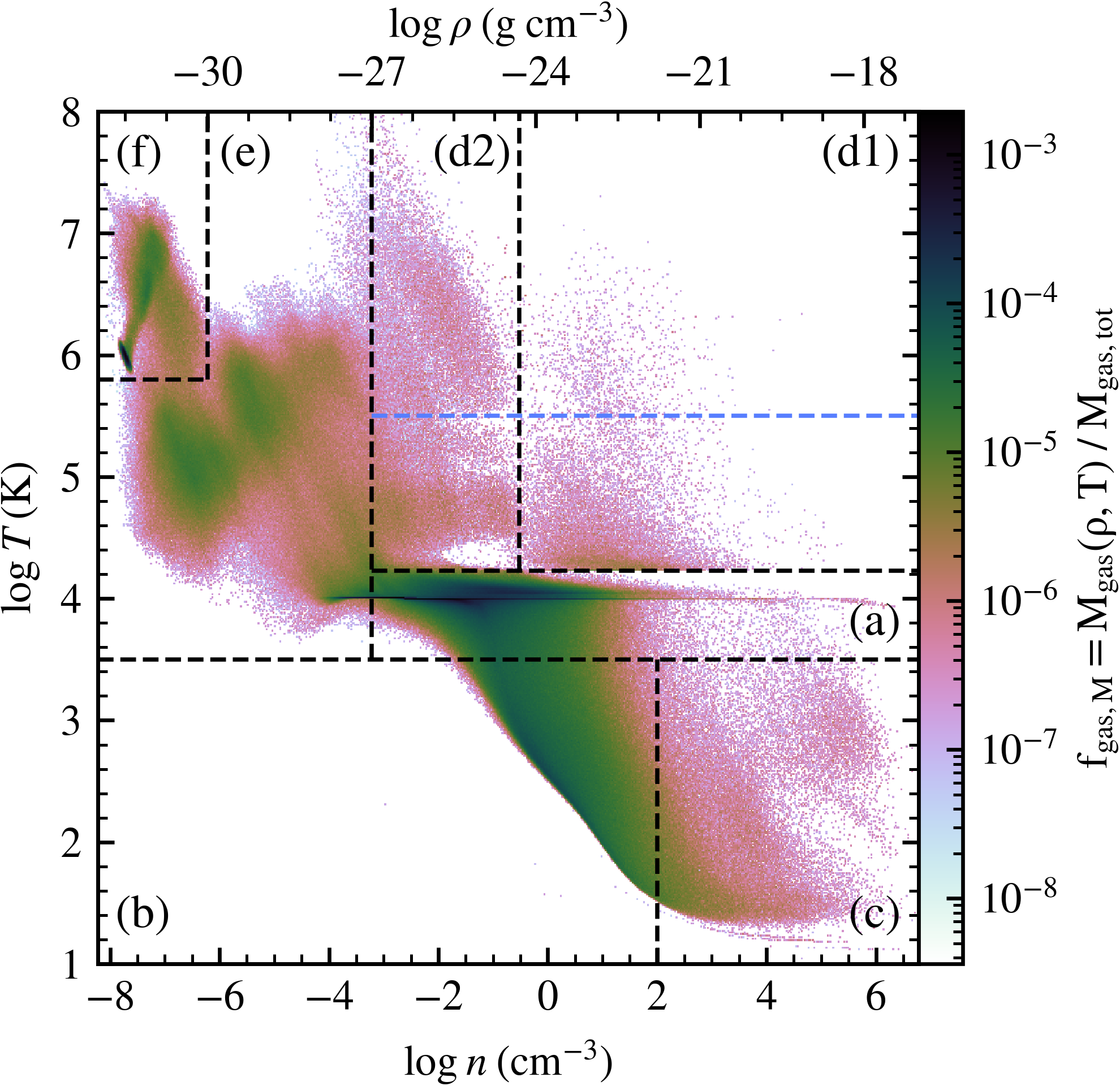}
    \caption{Phase diagram of gas for the SFE1 run at 0.4 Gyr in the form of a two-dimensional histogram in density and temperature space, with the color of each bin represents the mass fraction of gas at the corresponding density-temperature phase. Different phases are presented as different regions separated by dashed black lines.}
    \label{fig:phase}
\end{figure}

As mentioned in \citet{2019MNRAS.489.4233M}, the SMUGGLE model is capable to self-consistently generate a multi-phase structure of the ISM. Fig.~\ref{fig:phase} shows the phase diagram of temperature and density in SFE1 run at 0.4 Gyr.
The diagram is divided into multiple regions, largely following the approach of \citet{2019MNRAS.489.4233M}, with some necessary modifications made according to our study of superbubbles. 
We consider the interstellar medium (ISM) as the gas within the galactic disk, corresponding to phase (a) to (d). Most of the gas retains a warm state as phase (a), with a temperature of about $10^{4} ~\mathrm{K}$. Some of the gas will cool down into cold phase, collapse into dense clumps and form giant molecular clouds (GMCs) in phase (b), with some molecular cores evolve dense enough (i.e., $n > 100 \, \rm{} cm^{-3}$) and become star forming gas in phase (c). Phase (d1) and (d2) both represent the gas emerging from stellar feedback, with (d1) resulting mainly from radiative and stellar wind feedback while (d2) is dominated by supernova feedback. After stars are formed, radiation from stars may ionize the surrounding gas as early feedback, heat the gas to a temperature of about $1.7 \times 10^4 ~\mathrm{K}$ and push the gas away through radiation pressure and stellar winds. This gas primarily occupies phase (d1). The gas in phase (d2) will be even hotter and more diffuse due to supernova events. In this work, we mainly focus on the superbubbles, which largely correspond to the region (d1) and (d2) with temperatures higher than $10^{5.5} \; \rm K$ (indicated by the dashed blue line). We note that the warm gas (with temperature lower than $10^{5.5} \; \rm K$) in phase (d1) and (d2) largely corresponds to the observed \ion{H}{II} regions. 

Gas in phase (d2) may eventually flow out of the galaxy and into the circumgalactic medium (CGM), which is largely referred to as phase (e). The outflowing gas will undergo an expansion process, gradually decreasing the temperature and becoming more diffuse. Some of the gas in phase (e) is possible to fall back into the ISM as phase (a), and this kind of falling gas is probably the origin of the observed high velocity clouds \citep[HVCs, e.g.][]{1980ApJ...236..577B, 2010ApJ...709L.138L}. Intense stellar feedback in the galactic center can occasionally generate a massive outflow and inject the gas to distances comparable to, or even larger than, the virial radius. Some of the outflow gas may become too distant from the galaxy and cannot fall back, instead it will gradually mix with the gas in phase (f), which represents the background gas environment.
It is important to note that the simulations we employ do not account for the cosmological environment and there is no gas inflow from the IGM, so the CGM consists of only the outflow gas from the ISM and it might be considered as a lower limit to the presence of hot and diffuse gas.

%%%%%%%%%%%%%%%%%%%%%%%%%%%%%%%%%%%%%%%%%%%%%%%%%%

% Don't change these lines
\bsp	% typesetting comment
\label{lastpage}
\end{document}